\documentclass[aps,twocolumn,nofootinbib,showpacs,longbibliography]{revtex4-1}
\usepackage{bm,amsmath,amssymb,graphicx}
\usepackage[colorlinks,linkcolor=blue,urlcolor=blue]{hyperref}
\begin{document}

\title{A dynamically constrained Yang-Mills theory with Lorentz symmetry group as an alternative theory of gravity}

\author{Hans Christian \"Ottinger}
\email[]{hco@mat.ethz.ch}
\homepage[]{www.polyphys.mat.ethz.ch}
\affiliation{ETH Z\"urich, Department of Materials, CH-8093 Z\"urich, Switzerland}

\date{\today}

\begin{abstract}
We develop the complete composite theory of gravity, in which the gauge vector fields of the Yang-Mills theory with Lorentz symmetry group are expressed in terms of the tetrad variables obtained from the decomposition of a metric. A key element of a compelling formulation of composite gravity are refined coordinate conditions that offer a natural coupling of the gravitational field to matter and ensure the closest relationship to general relativity. The composite theory of gravity is presented from three different perspectives highlighting its intuitive interpretation, its relationship to general relativity and its canonical Hamiltonian formulation, where the latter clarifies the structure of the heavily constrained theory and provides the starting point for its quantization. The main physical ingredient of the theory is an anisotropic velocity-momentum relation, or tensorial mass, described by a metric. We discuss the static isotropic solution in great detail because it provides the background for the high-precision tests to be passed by an alternative theory of gravity and for the understanding of black holes.
\end{abstract}


\maketitle

\section{Introduction}
In view of its stunning mathematical elegance and its impressive physical predictions, Einstein's theory of gravity arguably is the most celebrated theory in physics. General relativity has shaped our thinking about space and time to an extent that it has become difficult to recollect the awareness that every physical theory comes with limitations. For a theory of gravity it is particularly natural to expect that lower and upper limits for the range of mass and length scales are vital to ensure its validity.

Newton's theory of gravity was developed for length scales ranging from the size of apple trees to the radius of our planetary system. In Cavendish's torsion balance for measuring Newton's constant (or ``weighing the world''), two lead balls of $158\,{\rm kg}$ and $0.73\,{\rm kg}$ were separated by $23\,{\rm cm}$. High-precision tests of Einstein's theory of general relativity involve length scales ranging from the radius of planetary orbits (anomalous precession of the perihelion of Mercury)  to the radius of the sun (deflection of light by the sun), smaller stars (gravitational redshifts of spectral lines from white dwarfs) or our planet (gravitational redshifts using terrestrial sources). Black hole observations probe strong gravitational interactions at the Schwarzschild radius, which is a few or a few tens of kilometers for stellar black holes.

But why should general relativity hold all the way down to length scales of the order of the Schwarzschild radius associated with elementary particles, which is the Planck length, or of the order of the size of the entire universe? Still unsurmountable problems with the quantization of general relativity and the need for dark energy to explain the accelerated expansion of the universe within Einstein's theory might indicate problems at both ends of the range of length and mass scales. These enormous challenges provide a strong motivation for proposing and investigating alternative theories of gravity.

The alternative theory of gravity developed in this paper is motivated by the long-standing challenge of the quantization of general relativity. As all other present theories of interactions between elementary particles, it is based on a gauge theory of the Yang-Mills type \cite{YangMills54}. The most appealing gauge symmetry group for a theory of gravity is the Lorentz group. Actually, immediately after the original work of Yang and Mills, Utiyama \cite{Utiyama56} considered the Yang-Mills theory based on the Lorentz group as a potential theory of gravity. The ideas of this pioneering work, which may be considered as the origin of what is now known as gauge gravitation theory \cite{CapozzielloDeLau11,IvanenkoSar83}, have been criticized as ``unnatural'' by Yang (see footnote~5 of \cite{Yang74}). Yang's own, allegedly more natural proposal \cite{Yang74} has itself been criticized massively in Chapter~19 of \cite{BlagojevicHehl}. While the construction of a theory of gravity from the Yang-Mills theory based on the Lorentz group clearly is a subtle matter, the potential for unifying all interactions and quantizing gravity suggests to continue the search for a more sophisticated construction within this elegant framework.

The Lorentz group has a six-dimensional Lie algebra and hence leads to a Yang-Mills theory with six gauge vector fields, so that it has way too many degrees of freedom for a theory of gravity. Therefore, we need a guiding principle for selecting a subset of solutions or, in other words, for introducing constraints. In the present approach we achieve a severe reduction of degrees of freedom by means of the framework of composite theories \cite{hco235,hco237}. The basic idea is to write the gauge vector fields of the Yang-Mills theory in terms of fewer, more fundamental variables and their derivatives. The admission of derivatives in this so-called composition rule implies that the composite theory involves higher than second derivatives. The power of the tool of composite theories stems from the fact that, in their Hamiltonian formulations \cite{hco235,hco237}, the structure of the constraints providing the selection principle is nicely transparent. The use of composite theories is a major advantage compared to previous investigations \cite{Utiyama56,Yang74} of the Yang-Mills theory based on the Lorentz group in the context of gravity.

As the composite theory of gravity \cite{hco231}, just like the underlying Yang-Mills theory, is formulated in a background Minkowski space-time, there arises the problem of how to characterize the ``good'' coordinate systems in which the theory may be applied. This characterization should be Lorentz invariant, but not invariant under more general coordinate transformations, that is, it shares the formal properties of coordinate conditions in general relativity. However, the unique solutions obtainable from Einstein's field equations only after specifying coordinate conditions are all physically equivalent, whereas the coordinate conditions in composite gravity characterize the physically preferred systems. From a historical perspective, it is remarkable that Einstein in 1914 still believed that the metric should be completely determined by the field equations and, therefore, a generally covariant theory of gravity was not desirable (see \cite{Giovanelli21} for a detailed discussion). The important task of characterizing the preferred systems in composite gravity is addressed in great detail in the present paper.

Compared to the original papers on the composite theory of gravity \cite{hco231,hco240}, the present work contains several important new developments. Most notably, the development of more appropriate coordinate conditions allows us to choose the coupling constant of the Yang-Mills theory such that its gauge vector fields correspond to a torsion-free connection, thus making the composite theory most similar to general relativity. Moreover, these coordinate conditions suggest a natural mechanism for coupling the gravitational field to matter. Finally, the correspondingly modified static isotropic solution automatically reproduces the high-precision predictions of general relativity. The combination of a gauge theory with the Lorentz group as its symmetry group and the framework of composite theories for introducing constraints does not only allow us to keep some of the mathematical elegance of general relativity, but also reproduces its most precisely tested physical predictions.

Although the so-called ``double-copy approach'' \cite{Bernetal08,Bernetal10} also tries to establish a relationship between Yang-Mills theories and theories of gravity and it has even been pointed out that its formulas ``hint at some notion of compositeness, albeit with a rather novel structure'' (see last paragraph of \cite{Bernetal10}), that approach is very different from the composite theory of gravity studied in the present paper. The double-copy approach is based on the observation that, in a suitable representation, there is an analogy between kinetic factors and color factors associated with the diagrams of perturbation theory. It starts from Yang-Mills theories with symmetry group SU(N) rather than the Lorentz group and relies on conjectures for the structure of perturbation theory rather than the general tool of composite theories \cite{hco235,hco237}.

The structure of the paper is as follows. In Sec.~\ref{secoverview}, we present and motivate the basic equations of the composite theory of gravity in their most intuitive and illuminating form; special emphasis is on a new type of coordinate conditions, and anisotropic velocity-momentum relations are identified as a core element of the theory. In Sec.~\ref{secmetricform}, we reformulate the theory in terms of metric, connection and curvature, thus clarifying the relationship to general relativity. In Sec.~\ref{secHamiltonian}, the canonical Hamiltonian formulation of composite theories offers still another perspective on composite gravity, clarifying the role of constraints and providing a solid starting point for quantization. In Sec.~\ref{secisosol}, we provide the equations characterizing the static isotropic solution of composite gravity from the different perspectives and solve them by Robertson expansions and numerical methods. Finally, we summarize and discuss our results.

\section{Overview of theory}\label{secoverview}
For a first overview of an alternative theory of gravity constructed from the Yang-Mills theory based on the Lorentz group, we focus on the field equations rather than on an underlying Lagrangian or Hamiltonian structure. We simply compile a complete set of evolution equations in the general spirit of composite theories. A particularly interesting feature of the proposed equations is the coupling mechanism between gravitational field and matter, which differs from a previous mechanism developed in the Lagrangian setting \cite{hco231} and in the Hamiltonian approach for the weak-field approximation \cite{hco240}.

\subsection{Yang-Mills theory based on Lorentz group}
Yang-Mills theories are characterized by Lorentz covariant field equations with additional continuous symmetries in a background Minkowski space-time. We denote the Minkowski metric with signature $(-,+,+,+)$ by $\eta_{\kappa\lambda} = \eta^{\kappa\lambda}$. Throughout this paper, the Minkowski metric is used for raising or lowering space-time indices.

The basic vector fields $A_{a \nu}$ of a Yang-Mills theory are labeled by two indices. In addition to the space-time index $\nu$, there is an index $a$ that labels a set of base vectors of the Lie algebra associated with an underlying continuous symmetry group. Yang-Mills theories are usually considered for the compact special unitary group ${\rm SU}(N)$ of unitary complex $N \times N$ matrices with determinant $1$ and the corresponding Lie algebra ${\rm su}(N)$, which is a vector space of dimension $N_{\rm p}=N^2-1$. In other words, the index $a$ takes values from $1$ to $N_{\rm p}$. Note that $N_{\rm p}$ may also be regarded as the number of continuous parameters required to characterize the elements of the underlying Lie group.

The elegance of the Yang-Mills equations for the vector fields $A_{a \nu}$ stems from the fact that they are covariant not only under Lorentz transformations (associated with the label $\nu$ of the fields) but also under infinitesimal local symmetry transformations from the underlying group (associated with the label $a$ of the fields),
\begin{equation}\label{gaugeA}
   \delta A_{a \rho} = \frac{\partial \Lambda_a}{\partial x^\rho}
   + \tilde{g} f^{bc}_a \, A_{b \rho} \, \Lambda_c ,
\end{equation}
where $\tilde{g}$ is a coupling constant and $f^{bc}_a$ stands for the structure constants of the Lie algebra representing the infinitesimal symmetry transformations characterized by the $N_{\rm p}$ functions $\Lambda_a$. These local symmetry transformations are often referred to as gauge transformations, and the two-label fields $A_{a \nu}$ as gauge vector fields. Gauge symmetry implies that only half of the degrees of freedom are physical. For example, the transverse modes of electromagnetic fields propagating through empty space are physically relevant, whereas the longitudinal and temporal modes lack direct physical significance. Very generally, one needs to find evolution equations for the physical degrees of freedom and ways of fixing the unphysical ones. The presence of unphysical degrees of freedom moreover raises deep questions about their proper treatment in quantization procedures. For gauge constraints, these issues can be handled elegantly by the BRST methodology (the acronym derives from the names of the authors of the original papers \cite{BecchiRouetStora76,Tyutin75}; see also \cite{Nemeschanskyetal86,hco229}).

For developing the composite theory of gravity \cite{hco231}, we depart from the Yang-Mills theory based on the Lorentz group ${\rm SO}(1,3)$ with Lie algebra ${\rm so}(1,3)$. This group consists of the real $4 \times 4$ matrices that leave the Minkowski metric invariant, that is, of the matrices representing rotations in space and Lorentz boosts mixing space and time. The total number of parameters for the Lorentz group is $N_{\rm p}=6$, three for characterizing rotations and three for the boosts. The following choice of the six base vectors of the Lie algebra is quite natural: three generators of boosts in the three spatial directions ($a=1,2,3$) and three generators of rotations around the three coordinate axes ($a=4,5,6$). More illuminating than the labels $a$ are the corresponding pairs of space-time indices $(\kappa,\lambda)$ given in Table~\ref{tabindexmatch}, which characterize the two-dimensional subspaces in which the corresponding generators act. Note that the boosts lead from the compact group of rotations ${\rm SO}(3)$ to the noncompact Lorentz group ${\rm SO}(1,3)$.

\begin{table}
\begin{tabular}{c|c c c c c c}
    $a$ \, & \, $1$ & $2$ & $3$ & $4$ & $5$ & $6$ \\
	\hline
	$(\kappa,\lambda)$ \,
    & \, $(0,1)$ & $(0,2)$ & $(0,3)$ & $(2,3)$ & $(3,1)$ & $(1,2)$ \\
\end{tabular}
\caption{Correspondence between the label $a$ for the base vectors of the six-dimensional Lie algebra ${\rm so}(1,3)$ of the Lorentz group and ordered pairs $(\kappa,\lambda)$ of space-time indices.}
\label{tabindexmatch}
\end{table}

Given any six components $\Lambda_a$, the quantities $\Lambda_{(\kappa,\lambda)}$ are directly defined for the index pairs given in Table~\ref{tabindexmatch}. It is natural to define $\Lambda_{(\kappa,\lambda)}$ for all pairs of space-time indices by antisymmetric continuation. The antisymmetric matrix  $\Lambda_{(\kappa,\lambda)}$ with six degrees of freedom contains no more and no less information than the original $\Lambda_a$. Table~\ref{tabindexmatch} specifies how to fill up the antisymmetric matrix $\Lambda_{(\kappa,\lambda)}$ with the elements $\pm\Lambda_a$, or how to extract $\Lambda_a$ from $\Lambda_{(\kappa,\lambda)}$.

Following standard procedures for Yang-Mills theories (see, e.g., Sect.~15.2 of \cite{PeskinSchroeder}, Chap.~15 of \cite{WeinbergQFT2}, or \cite{hco229}), we can introduce a field tensor in terms of the gauge vector fields,
\begin{equation}\label{Fdefinition}
   F_{a \mu\nu} = \frac{\partial A_{a \nu}}{\partial x^\mu}
   - \frac{\partial A_{a \mu}}{\partial x^\nu}
   + \tilde{g} f^{bc}_a A_{b \mu} A_{c \nu} ,
\end{equation}
where $f^{bc}_a$ now stands for the structure constants of the Lorentz group. A Lie algebra label, say $a$, can be raised or lowered by raising or lowering the indices of the pairs associated with $a$ according to Table~\ref{tabindexmatch} by means of the Minkowski metric. The structure constants can then be specified as follows: $f^{abc}$ is $1$ ($-1$) if $(a,b,c)$ is an even (odd) permutation of $(4,5,6)$, $(1,3,5)$, $(1,6,2)$ or $(2,4,3)$, and $0$ otherwise. An occasionally useful, explicit formula for $f^{abc}$ can be found in Eq.~(\ref{Lorentzstructure}) of Appendix~\ref{Appcovdevs}. By means of the Jacobi identity for the structure constants,
\begin{equation}\label{Jacobiid}
   f^{sb}_a f^{cd}_s + f^{sc}_a f^{db}_s + f^{sd}_a f^{bc}_s = 0 ,
\end{equation}
the gauge transformation behavior of the field tensor is found to be
\begin{equation}\label{gaugeF}
   \delta F_{a \mu\nu} = \tilde{g} f^{bc}_a \, F_{b \mu\nu} \, \Lambda_c .
\end{equation}

In terms of the field tensor $F_{a \mu\nu}$, the field equations of Yang-Mills theory can be written as
\begin{equation}\label{YMfieldeqs}
   \frac{\partial F_{a \mu\nu}}{\partial x_\mu}
   + \tilde{g} f_a^{bc} \, A_b^\mu F_{c \mu\nu} = - J_{a \nu} = 0 ,
\end{equation}
where we have assumed that the external sources $J_{a \nu}$ vanish. Even in the presence of matter, we assume the external sources to be negligible. The gauge transformation laws (\ref{gaugeA}) and (\ref{gaugeF}) imply
\begin{eqnarray}
   \delta \left( \frac{\partial F_{a \mu\nu}}{\partial x_\mu}
   + \tilde{g} f_a^{bc} \, A_b^\mu F_{c \mu\nu} \right) &=& \nonumber\\
   && \hspace{-7em} \tilde{g} f_a^{bc} \left( \frac{\partial F_{b \mu\nu}}{\partial x_\mu}
   + \tilde{g} f_b^{de} \, A_d^\mu F_{e \mu\nu} \right) \Lambda_c . \qquad 
\label{gaugefieldeqs}
\end{eqnarray}
This result demonstrates the gauge invariance of the Yang-Mills field equations (\ref{YMfieldeqs}), which is the desired continuous symmetry of the theory.

The field equations (\ref{YMfieldeqs}) provide second-order differential equations for the gauge vector fields $A_{a \nu}$. Note that the second derivative terms in the field equations, that is,
\begin{equation}\label{YMfieldeqs2}
   \square A_{a \nu} - \frac{\partial}{\partial x^\nu} \frac{\partial A_{a \mu}}{\partial x_\mu} ,
\end{equation}
with the d'Alembertian second-order differential operator
\begin{equation}\label{squaredef}
   \square = \frac{\partial^2}{\partial x^\mu \partial x_\mu} =
   - \frac{\partial^2}{c^2 \, \partial t^2} + \Delta ,
\end{equation}
where $\Delta$ is the Laplace operator, provide the second time derivatives of $A_{a \nu}$ for $\nu > 0$, but not for $\nu = 0$. The second-derivative terms (\ref{YMfieldeqs2}) take a particularly simple form and the second time derivatives of all $A_{a \nu}$ are determined by the field equations if we impose the covariant Lorenz gauge condition
\begin{equation}\label{Lorenzgauge}
   \frac{\partial A_{a \mu}}{\partial x_\mu}  = 0 .
\end{equation}

The Yang-Mills theory based on the Lorentz group is clearly not suited as a theory of gravity because it possesses way too many degrees of freedom. It involves six four-vector fields satisfying second-order differential equations, which amounts to $48$ degrees of freedom (that is, $48$ initial conditions). Even if this number is reduced to $24$ by gauge conditions, including the Lorenz gauge (\ref{Lorenzgauge}) relating temporal and longitudinal field components, we are still left with too many degrees of freedom. The idea of composite gravity is to select a small subset of the solutions of the Yang-Mills theory by expressing the gauge vector fields in terms of fewer, more basic fields, as elaborated in the next step.

\subsection{Composition rule}
The basic idea of composite gravity is is to express the $24$ components of the gauge vector fields in terms of the $16$ fields ${b^\kappa}_\mu$ and their derivatives,
\begin{eqnarray}
   A_{(\kappa\lambda) \rho} &=&
   \frac{1}{2} \, \mbox{$\bar{b}^\mu$}_{\kappa} \left( \frac{\partial g_{\nu\rho}}{\partial x^\mu}
   - \frac{\partial g_{\mu\rho}}{\partial x^\nu} \right) \mbox{$\bar{b}^\nu$}_{\lambda}
   \nonumber\\
   &+& \frac{1}{2 \tilde{g}} 
   \left(\frac{\partial b_{\kappa\mu}}{\partial x^\rho} \, \mbox{$\bar{b}^\mu$}_{\lambda}
   - \mbox{$\bar{b}^\mu$}_{\kappa} \, \frac{\partial b_{\lambda\mu}}{\partial x^\rho} 
   \right) , \qquad 
\label{compositionrule}
\end{eqnarray}
where $\mbox{$\bar{b}^\mu$}_{\kappa}$ is the inverse of the regular matrix ${b^\kappa}_\mu$ and
\begin{equation}\label{localinertialdecomp}
   g_{\mu\nu} = \eta_{\kappa\lambda} \, {b^\kappa}_\mu {b^\lambda}_\nu
   = {b^\kappa}_\mu \, b_{\kappa\nu} .
\end{equation}
The gauge degrees of freedom arise from the fact that we consider only the matrix $g_{\mu\nu}$ as physical, not the individual factors ${b^\kappa}_\mu$ in the decomposition (\ref{localinertialdecomp}). The invariance properties of the Minkowski metric imply that ${b^\kappa}_\mu$ can be multiplied from the left with an arbitrary Lorentz transformation matrix without changing $g_{\mu\nu}$. This observation reveals the origin of the Lorentz symmetry group in composite gravity. Infinitesimal gauge transformations can be expressed in the form
\begin{equation}\label{gaugeb}
   \delta b_{\kappa\mu} = \tilde{g} \, \Lambda_{(\kappa\lambda)} {b^\lambda}_\mu ,
\end{equation}
where $\Lambda_{(\kappa\lambda)}$ is an arbitrary antisymmetric $4 \times 4$ matrix, which is equivalent to the six quantities $\Lambda_a$. It has been shown in \cite{hco231} that Eq.~(\ref{gaugeb}) leads to the gauge transformation behavior (\ref{gaugeA}) of the gauge vector fields defined in Eq.~(\ref{compositionrule}). The occurrence of derivatives in the composition rule (\ref{compositionrule}) is required for obtaining the contribution $\partial \Lambda_a/\partial x^\rho$ in the transformation law (\ref{gaugeA}).

Equation (\ref{localinertialdecomp}) may be regarded as a local transformation of the Minkowski metric to the background coordinate system. Therefore, the resulting matrix $g_{\mu\nu}$ has all the properties required for a metric in general relativity, and the quantities ${b^\kappa}_\mu$ can be regarded as tetrad or \emph{vierbein} variables. More intuitively, the tetrad variables provide a relation between the underlying Minkowski coordinate system and freely falling local coordinate systems. However, the metric cannot be used in the same fully geometric way as in general relativity because the Yang-Mills theory is defined in a fixed background Minkowski space-time. The proper physical interpretation of the metric $g_{\mu\nu}$ in composite gravity is revealed in Sec.~\ref{secparticlemotion}.

The two-fold role of the Lorentz group in composite gravity can be recognized nicely in the context of the tetrad variables ${b^\kappa}_\mu$. Lorentz transformations can independently act on the space-time indices $\kappa$ and $\mu$. On the one hand, when a Lorentz transformation acts on the index $\kappa$, it represents a local symmetry transformation leaving the metric (\ref{localinertialdecomp}) invariant. On the other hand, when a Lorentz transformation acts on the index $\mu$, it represents a global change of the coordinate system in the underlying Minkowski space. The index $\mu$ makes ${b^\kappa}_\mu$ a set of Lorentz four-vector fields, the index $\kappa$ introduces internal degrees of freedom associated with the Lorentz group as symmetry group.

\subsection{Coordinate conditions}\label{seccoco}
Closer inspection of the composition rule (\ref{compositionrule}) (see Sec.~\ref{secHamconstraintscomp} for details) reveals that it consists of $12$ evolution equations and $12$ constraints for the tetrad variables. The composition rule does not provide the time derivatives of the components $g_{0\mu}$. We need four more evolution equations for $g_{0\mu}$ to arrive at a complete set of $16$ evolution equations for the $16$ tetrad variables.

By inserting the composition rule (\ref{compositionrule}) into the Yang-Mills field equations (\ref{YMfieldeqs}) we obtain third-order partial differential equations for ${b^\kappa}_\mu$. We hence formulate also the missing evolution equations as Lorentz covariant third-order differential equations,
\begin{equation}\label{coco3mat}
   \square \left( \frac{\partial g_{\mu\rho}}{\partial x_\rho}
   - \frac{1}{2} \frac{\partial {g_\rho}^\rho}{\partial x^\mu} \right)
   + \frac{\Lambda}{2} \frac{\partial {g_\rho}^\rho}{\partial x^\mu} =
   \frac{4 \pi G}{c^4} \, \frac{\partial {T_\rho}^\rho}{\partial x^\mu} .
\end{equation}
These equations have the status of coordinate conditions, that is, they characterize the background Minkowski systems for which the composite theory of gravity is meant to be applicable. This is a crucial ingredient for any theory without general covariance. The term in parentheses on the left-hand side corresponds to the harmonic coordinate conditions of linearized general relativity (see, e.g., Eq.~(10.1.9) of \cite{Weinberg}). A non-vanishing right-hand side, where $c$ is the speed of light, $G = 6.674 \cdot 10^{-11} \, {\rm m}^3 / ({\rm kg} \, {\rm s}^2)$ is Newton's constant and ${T_\mu}^\nu$ is the energy-momentum tensor of matter, is required for reasons of consistency with Newton's theory of gravitation. Finally, if only to avoid later regrets, we have introduced a cosmological term with the cosmological constant $\Lambda$. Note that the right-hand side provides the only coupling between gravitational field and matter in composite gravity, which is taken to be of a most rudimentary scalar form. The linear form of the left-hand side is not just chosen for simplicity. If we wish composite gravity to be independent of an overall scale factor in the metric, the linear form is suggested by the fact that the energy-momentum tensor on the right-hand side turns out to be also linear in the metric (see Eq.~(\ref{particleT}) below).

If we act on the coordinate conditions with $\partial/\partial x_\mu$ and sum over $\mu$, we obtain the fourth-order differential equation that results from acting with the d'Alembertian $\square$ on the equation,
\begin{equation}\label{coco3mati}
   \frac{\partial^2 g_{\rho\sigma}}{\partial x_\rho\partial x_\sigma} -
   \frac{1}{2} \, \square \, {g_\rho}^\rho
   + \frac{\Lambda}{2} \, {g_\rho}^\rho = \frac{4 \pi G}{c^4} \, {T_\rho}^\rho .
\end{equation}
It seems natural to assume that this scalar second-order condition must be satisfied as a consistency requirement coming with the coordinate conditions (\ref{coco3mat}) (as a result of trivial boundary conditions at infinity). Equation~(\ref{coco3mati}) reproduces Newton's law $\nabla^2 g_{00} = - 8 \pi G T_{00}/c^4$ for weak static fields generated by nonrelativistic matter and $\Lambda=0$ (see, e.g., Eq.~(7.1.3) of \cite{Weinberg}), that is, when metric and energy-momentum tensor are dominated by $g_{00}$ and $T_{00}$, respectively, and all time derivatives vanish.

By using Eq.~(\ref{coco3mati}) to eliminate the energy-momentum tensor from Eq.~(\ref{coco3mat}), we obtain the coordinate conditions in the alternative form
\begin{equation}\label{coco3mat0}
   \square  \frac{\partial g_{\mu\rho}}{\partial x_\rho} - \frac{\partial}{\partial x^\mu}
   \frac{\partial^2 g_{\rho\sigma}}{\partial x_\rho\partial x_\sigma} = 0 .
\end{equation}
Given Eq.~(\ref{coco3mati}), the conditions (\ref{coco3mat}) and (\ref{coco3mat0}) are equivalent. If we choose the form  (\ref{coco3mat0}) of the coordinate conditions, the only coupling of the gravitational field to matter arises from the required additional second-order differential equation (\ref{coco3mati}).

\subsection{Particle motion in gravitational field}\label{secparticlemotion}
In the development of the field equations of composite gravity, the metric appeared as an auxiliary variable without immediate physical meaning. Through the coordinate conditions (\ref{coco3mat}), the metric gains physical significance by being related to matter. However, we still need to provide a direct interpretation of $g_{\mu\nu}$ and the effect of gravity on matter.

The basic physical significance of the metric $g_{\mu\nu}$ is that it provides an anisotropic relationship between velocity and momentum. This minimalist assumption has far-reaching consequences, as we show now.

Let us consider a point particle of rest mass $m$ moving along a trajectory $\bar{x}^\mu(t)$ in the underlying Minkowski space-time. Then, the anisotropic velocity-momentum relation is given by
\begin{equation}\label{velmomrel}
   p_\mu = \gamma m \, g_{\mu\nu} \frac{d\bar{x}^\nu}{d t} ,
\end{equation}
and the energy-momentum (flux) tensor of matter can be written as (see, e.g., Eq.~(2.8.2) of \cite{Weinberg})
\begin{equation}\label{particleT}
   {T_\mu}^\nu = \gamma m \, g_{\mu\sigma} \frac{d\bar{x}^\sigma}{d t}
   \frac{d\bar{x}^\nu}{d t} \, \delta^3(\bm{x}-\bar{\bm{x}}(t)) ,
\end{equation}
where the Lorentz factor $\gamma$ in the presence of gravity is defined as
\begin{equation}\label{gammadef}
   \gamma = c \left( - g_{\mu\nu} \frac{d\bar{x}^\mu}{d t} \frac{d\bar{x}^\nu}{d t}
   \right)^{-\frac{1}{2}} .
\end{equation}
The physical reason for the appearance of $g_{\mu\nu}$ becomes clear in both Eq.~(\ref{particleT}) and Eq.~(\ref{gammadef}). In the energy-momentum tensor (\ref{particleT}), the lower index $\mu$ is associated with momentum and the upper index $\nu$ with flow. Whereas flow or convection is properly characterized by the particle velocity, momentum requires the extra factors of $m$ and $g_{\mu\sigma}$ accomplishing the velocity-momentum relation. As $T_{00}$ is the energy density, the occurrence of momentum times velocity is recognized as more natural in the energy than the familiar square of either velocity or momentum in the kinetic energy. The same is assumed to be true for the square of velocity in the Lorentz factor, which is reflected in the occurrence of $g_{\mu\nu}$ in the definition (\ref{gammadef}).

Equations (\ref{particleT}) and (\ref{gammadef}) imply the following result for the trace of the energy-momentum tensor for a point mass, 
\begin{equation}\label{particleTtr}
   {T_\rho}^\rho = - \frac{m c^2}{\gamma} \, \delta^3(\bm{x}-\bar{\bm{x}}(t)) .
\end{equation}
This expression is to be used as the source term in the coordinate conditions (\ref{coco3mat}) and in the related second-order differential equation (\ref{coco3mati}). Energy-momentum conservation is obtained in the form
\begin{equation}\label{particleTdiv}
   \frac{\partial{T_\mu}^\nu}{\partial x^\nu} =
   \frac{d p_\mu}{d t} \, \delta^3(\bm{x}-\bar{\bm{x}}(t)) .
\end{equation}

As the Hamiltonian for the point particle is given by $H_{\rm m}=-c p_0$, we are now in a position to formulate the equation of motion for the particle. From Eqs.~(\ref{velmomrel}) and (\ref{gammadef}) we obtain
\begin{equation}\label{particleppm}
    - \bar{g}^{\mu\nu} p_\mu p_\nu = m^2 c^2 ,
\end{equation}
where $\bar{g}^{\mu\nu}$ is the inverse of the metric matrix $g_{\mu\nu}$, evaluated at $x^\mu = (ct, \bar{\bm{x}}(t))$. By solving this quadratic equation for $p_0$, we obtain the Hamiltonian $H_{\rm m}$ in terms of the momenta $p_j$,
\begin{equation}\label{particleH}
   H_{\rm m} = c \sqrt{\frac{m^2 c^2 + \bar{g}^{ij} p_i p_j}{-\bar{g}^{00}}
   + \left( \frac{\bar{g}^{0j} p_j}{\bar{g}^{00}}\right)^2}
   + c \frac{\bar{g}^{0j} p_j}{\bar{g}^{00}} .
\end{equation}
This Hamiltonian for a point mass [after considerable rearrangements or according to Eq.~(\ref{dHmdp2})] leads to the evolution equations
\begin{equation}\label{particlev3vec}
   \frac{d \bar{x}^j}{d t} = \frac{\partial H_{\rm m}}{\partial p_j}
   = \frac{1}{\gamma m} \, \bar{g}^{j\mu} p_\mu ,
\end{equation}
which is consistent with Eq.~(\ref{velmomrel}), and
\begin{equation}\label{particlea3vec}
   \frac{d p_j}{d t} = - \frac{\partial H_{\rm m}}{\partial \bar{x}^j}
   = - \frac{p_\mu p_\nu}{2 \gamma m} \frac{\partial \bar{g}^{\mu\nu}}{\partial \bar{x}^j}
   = \frac{\gamma m}{2} \frac{d \bar{x}^\mu}{d t} \frac{d \bar{x}^\nu}{d t}
   \frac{\partial g_{\mu\nu}}{\partial \bar{x}^j} .
\end{equation}
Equations (\ref{particlev3vec}) and (\ref{particlea3vec}) describe the cogeodesic flow resulting from the Hamiltonian approach. Equation~(\ref{velmomrel}) implies that Eq.~(\ref{particlev3vec}) holds also for $j=0$. The first part of Eq.~(\ref{particlea3vec}) for $j=0$ expresses the fact that the energy changes only through the explicit time dependence of the Hamiltonian. By differentiating  Eq.~(\ref{particleppm}) with respect to time and using Eqs.~(\ref{velmomrel}) and (\ref{particlea3vec}), we obtain
\begin{equation}\label{particleavec0}
   p_\mu p_\nu \frac{\partial \bar{g}^{\mu\nu}}{\partial t}
   = - 2 \gamma m c \, \frac{d p_0}{d t} ,
\end{equation}
which implies that the entire sequence of equations in Eq.~(\ref{particlea3vec}) holds also for $j=0$. By using Eq.~(\ref{velmomrel}), we can rewrite the equations of Hamiltonian cogeodesic flow in the more common form
\begin{equation}\label{particlegeodesic}
    \gamma \frac{d}{d t} \gamma \frac{d \bar{x}^\rho}{d t} + \Gamma^\rho_{\mu\nu} 
    \gamma^2 \, \frac{d \bar{x}^\mu}{d t} \frac{d \bar{x}^\nu}{d t} = 0 .
\end{equation}
Note that $\gamma d/dt$ is the derivative with respect to proper time and $\Gamma^\rho_{\mu\nu}$ is the Christoffel symbol associated with the metric $g_{\mu\nu}$. It is remarkable that the geodesic form of the equation of motion is a consequence of the anisotropic velocity-momentum relation. For geodesic particle motion, the energy-momentum balance (\ref{particleTdiv}) becomes
\begin{equation}\label{particleTdivg}
   \frac{\partial{T_\mu}^\nu}{\partial x^\nu} = \frac{1}{2} \bar{g}^{\rho\nu}
   \frac{\partial g_{\nu\sigma}}{\partial x^\mu} \, {T_\rho}^\sigma .
\end{equation}

In summary of our overview in Sec.~\ref{secoverview}, the composite theory of the gravitational field is given by the free Yang-Mills field equations (\ref{YMfieldeqs}) with the Lorentz group as underlying symmetry group, the Lorenz gauge condition (\ref{Lorenzgauge}), the composition rule (\ref{compositionrule}) for the gauge vector fields in terms of tetrad variables explaining the origin of the gauge symmetry, and the coordinate conditions (\ref{coco3mat}), where only the latter couple the gravitational field to matter. A minimalist interpretation of the metric as an anisotropic velocity-momentum relation is sufficient for obtaining the geodesic equation of motion for a point mass in a gravitational field. No deformation of the underlying Minkowski space-time is required.

\section{Evolution equations for metric}\label{secmetricform}
The idea of this section is to tell the story of composite gravity developed in the preceding section a second time, now from the perspective of general relativity. The decomposition (\ref{localinertialdecomp}) provides an algebraic relation between tetrad variables and metric. Considering first derivatives, we wish to establish a relation between the gauge vector fields (\ref{compositionrule}) and a connection associated with the metric. Proceeding to second derivatives, we look for a relation between the Yang-Mills field tensor (\ref{Fdefinition}) and the curvature tensor associated with the metric. The field equations of composite gravity can finally be rewritten as second-order differential equations for the connection or as third-order differential equations for the metric.

Structural correspondences between the Yang-Mills theory based on the Lorentz group and general relativity have previously been recognized in the work of Utiyama \cite{Utiyama56}. Even deeper relations with the Einstein-Cartan theory of gravity, which allows for geometries with torsion, have been revealed by Kibble \cite{Kibble61} and Sciama \cite{Sciama62ip}.

\subsection{Covariant derivatives, connection and curvature}
The subsequent developments can be simplified by transforming a given field $B_a$ labeled by a Lie algebra index to a quantity with two space-time indices,
\begin{equation}\label{Btildef}
   \tilde{B}_{\mu\nu} = {b^\kappa}_\mu {b^\lambda}_\nu \, B_{(\kappa\lambda)} ,
\end{equation}
where $B_{(\kappa\lambda)}$ is given by Table~\ref{tabindexmatch} combined with antisymmetric continuation. The antisymmetry $\tilde{B}_{\mu\nu}=-\tilde{B}_{\nu\mu}$ is inherited. The gauge transformation behavior of $\tilde{B}_{\mu\nu}$ is obtained from Eqs.~(\ref{gaugeb}) and (\ref{supauxf1}),
\begin{equation}\label{gaugeBtil}
   \delta \tilde{B}_{\mu\nu} = {b^\kappa}_\mu {b^\lambda}_\nu \left[
   \delta B_{(\kappa\lambda)} - \tilde{g} f^{bc}_{(\kappa\lambda)} \, B_b \, \Lambda_c \right] ,
\end{equation}
where the second term in square brackets typically simplifies the transformation behavior of $B_{(\kappa\lambda)}$ by canceling a term in $\delta B_{(\kappa\lambda)}$. We immediately realize how useful the definition (\ref{Btildef}) is because Eq.~(\ref{gaugeBtil}) implies that the gauge transformation law (\ref{gaugeF}) for the field tensor $F_{a \, \mu'\nu'}$ leads to a gauge invariant tensor $\tilde{F}_{\mu\nu \, \mu'\nu'}$. The definition (\ref{compositionrule}) of the gauge vector fields simplifies to
\begin{equation}\label{Atildef}
   \tilde{A}_{\mu\nu\rho} = \frac{1}{2} \left( \frac{\partial g_{\nu\rho}}{\partial x^\mu}
   - \frac{\partial g_{\mu\rho}}{\partial x^\nu} \right) + \frac{1}{2 \tilde{g}} 
   \left( {b^\kappa}_\mu \, \frac{\partial b_{\kappa\nu}}{\partial x^\rho}
   - \frac{\partial{b^\kappa}_\mu}{\partial x^\rho} \, b_{\kappa\nu} \right) ,  
\end{equation}
where the first term consists of derivatives of the metric and is gauge invariant. We can rewrite this equation as
\begin{equation}\label{Atildefx}
   \tilde{A}_{\mu\nu\rho} = \frac{1}{\tilde{g}} \left( 
   {b^\kappa}_\mu \, \frac{\partial b_{\kappa\nu}}{\partial x^\rho}
   - \tilde{\Gamma}_{\mu\rho\nu} \right) ,
\end{equation}
with
\begin{equation}\label{Gammabardef}
   \tilde{\Gamma}_{\sigma\mu\nu} = \frac{1}{2} 
   \left[ \frac{\partial g_{\sigma\nu}}{\partial x^\mu} + \tilde{g}
   \left( \frac{\partial g_{\mu\sigma}}{\partial x^\nu}
   - \frac{\partial g_{\mu\nu}}{\partial x^\sigma} \right) \right] ,
\end{equation}
which is closely related to the connection
\begin{equation}\label{Gammadef}
   \Gamma^\rho_{\mu\nu} = \bar{g}^{\rho\sigma} \, \tilde{\Gamma}_{\sigma\mu\nu} .
\end{equation}
Note that this connection $\Gamma^\rho_{\mu\nu}$ coincides with the Christoffel symbols occurring in the geodesic equation of motion (\ref{particlegeodesic}) only for $\tilde{g}=1$. For $\tilde{g} \ne 1$, $\Gamma^\rho_{\mu\nu}$ is not symmetric in $\mu$ and $\nu$. This lack of symmetry indicates the presence of torsion. Note, however, that the connection is metric-compatible for all $\tilde{g}$ \cite{Jimenezetal19}, that is,
\begin{equation}\label{metriccompatible}
   \frac{\partial g_{\mu\nu}}{\partial x^\rho}
   - \Gamma^\sigma_{\rho\mu} g_{\sigma\nu}
   - \Gamma^\sigma_{\rho\nu} g_{\mu\sigma} = 0 ,
\end{equation}
which can be recast in the convenient form
\begin{equation}\label{metriccompatiblex}
   \frac{\partial g_{\mu\nu}}{\partial x^\rho} =
   \tilde{\Gamma}_{\mu\rho\nu} + \tilde{\Gamma}_{\nu\rho\mu} .
\end{equation}

We are now in a position to formulate a deep relation between covariant derivatives associated with connections on the one hand and covariant derivatives associated with the Yang-Mills theory based on the Lorentz group on the other hand. From the structure constants of the Lorentz group we obtain the following beautiful relation highlighting the fundamental role of the transformation (\ref{Btildef}) (for a proof, see Appendix~\ref{Appcovdevs}),
\begin{eqnarray}
   \frac{\partial \tilde{B}_{\mu\nu}}{\partial x^\rho}
   - \Gamma^\sigma_{\rho\mu} \tilde{B}_{\sigma\nu}
   - \Gamma^\sigma_{\rho\nu} \tilde{B}_{\mu\sigma} &=& \nonumber\\
   && \hspace{-9em} {b^\kappa}_\mu {b^\lambda}_\nu \left[
   \frac{\partial B_{(\kappa\lambda)}}{\partial x^\rho}
   + \tilde{g} \, f_{(\kappa\lambda)}^{bc} A_{b\rho} B_c \right] . \qquad 
\label{central}
\end{eqnarray}

In Appendix~\ref{Appfieldtensor}, it is shown that the field tensor (\ref{Fdefinition}) can be written in the alternative form
\begin{eqnarray}
   \tilde{F}_{\mu\nu \mu'\nu'} &=& \nonumber\\
   && \hspace{-2em} 
   \frac{1}{2} \left( 
   \frac{\partial^2 g_{\nu\nu'}}{\partial x^\mu \partial x^{\mu'}}
   - \frac{\partial^2 g_{\nu\mu'}}{\partial x^\mu \partial x^{\nu'}}
   - \frac{\partial^2 g_{\mu\nu'}}{\partial x^\nu \partial x^{\mu'}}
   + \frac{\partial^2 g_{\mu\mu'}}{\partial x^\nu \partial x^{\nu'}} \right)
   \nonumber\\
   && \hspace{-2em} + \, \frac{1}{\tilde{g}} \, \bar{g}^{\rho\sigma}
   ( \tilde{\Gamma}_{\rho\mu'\mu} \tilde{\Gamma}_{\sigma\nu'\nu}
   - \tilde{\Gamma}_{\rho\nu'\mu} \tilde{\Gamma}_{\sigma\mu'\nu} ) .
\label{Fdefinitiontilz}
\end{eqnarray}
This explicit expression for $\tilde{F}_{\mu\nu \mu'\nu'}$ reveals its symmetry properties: antisymmetry under $\mu \leftrightarrow \nu$ and $\mu' \leftrightarrow \nu'$ and, more surprisingly, symmetry under $(\mu\nu) \leftrightarrow (\mu'\nu')$.

Note that the transformed field tensor (\ref{Fdefinitiontilz}) of the Yang-Mills theory is very similar to the Riemann curvature tensor (see, e.g.\ \cite{Weinberg} or \cite{Jimenezetal19}),
\begin{equation}\label{R4def}
   {R^\mu}_{\nu\mu'\nu'} = \frac{\partial \Gamma^\mu_{\mu'\nu}}{\partial x^{\nu'}}
   - \frac{\partial \Gamma^\mu_{\nu'\nu}}{\partial x^{\mu'}}
   + \Gamma^\sigma_{\mu'\nu} \Gamma^\mu_{\nu'\sigma}
   - \Gamma^\sigma_{\nu'\nu} \Gamma^\mu_{\mu'\sigma} .
\end{equation}
By means of Eqs.~(\ref{Gammadef}) and (\ref{metriccompatiblex}) we find the remarkable identity
\begin{equation}\label{Fdefinitiontily}
   \tilde{g} \, \bar{g}^{\mu\rho} \tilde{F}_{\rho\nu \mu'\nu'} = {R^\mu}_{\nu\mu'\nu'} ,
\end{equation}
which holds for all values of the coupling constant $\tilde{g}$.

The connection between composite gravity and general relativity becomes particularly close when we choose the coupling constant $\tilde{g}=1$. Then the connection (\ref{Gammabardef}), (\ref{Gammadef}) is torsion-free and given by the Christoffel symbols occurring also in the geodesic equation of motion (\ref{particlegeodesic}). For our previous choices of the coordinate conditions (quasi-Minkowskian coordinates in Section~V.A of \cite{hco231}; linearized harmonic coordinates for the weak-field approximation in Section~II.C of \cite{hco240}), the coupling constant $\tilde{g}=1$ was found to be inconsistent with the high-precision predictions of general relativity. With the new third-order coordinate conditions (\ref{coco3mat}), however, we have the big advantage that we can actually choose $\tilde{g}=1$, which we assume from now on.

\subsection{Alternative forms of field equations}\label{secfieldeqs}
With the help of Eq.~(\ref{central}), the standard field equations (\ref{YMfieldeqs}) for our Yang-Mills theory based on the Lorentz group can be written in the manifestly gauge invariant form
\begin{equation}\label{YMfieldeqsFtil}
   \eta^{\mu'\mu''} \left( \frac{\partial \tilde{F}_{\mu\nu \mu''\nu'}}{\partial x^{\mu'}}
   - \Gamma^\sigma_{\mu'\mu} \tilde{F}_{\sigma\nu \mu''\nu'}
   - \Gamma^\sigma_{\mu'\nu} \tilde{F}_{\mu\sigma \mu''\nu'} \right) = 0.
\end{equation}
By means of Eq.~(\ref{Fdefinitiontily}), these field equations can be rewritten in terms of the Riemann curvature tensor,
\begin{equation}\label{YMfieldeqsR}
   \eta^{\rho\nu'} \left( \frac{\partial {R^\mu}_{\nu\mu'\nu'}}{\partial x^\rho}
   + \Gamma^\mu_{\rho\sigma} {R^\sigma}_{\nu\mu'\nu'}
   - \Gamma^\sigma_{\rho\nu} {R^\mu}_{\sigma\mu'\nu'} \right) = 0.
\end{equation}
In view of Eq.~(\ref{R4def}), this latter form of the field equations is entirely in terms of the variables $\Gamma^\rho_{\mu\nu}$. However, this observation is not particularly useful as the coordinate conditions (\ref{coco3mat}) cannot be expressed in terms of $\Gamma^\rho_{\mu\nu}$ alone. The coordinate conditions, in general, keep us from a two-step procedure in which one first solves second-order differential equations for $\Gamma^\rho_{\mu\nu}$ and then, in a post-processing step, one obtains the metric from the differential equations (\ref{Gammabardef}) and (\ref{Gammadef}).

Therefore, we finally write the field equations directly as third-order differential equations for the metric. We write all third and second derivatives of the metric explicitly, whereas first derivatives are conveniently combined into connection variables. As the result we find the following set of equations for the composite theory of gravity obtained by expressing the gauge vector fields of the Yang-Mills theory based on the Lorentz group in terms of the tetrad variables obtained by factorizing the metric,
\begin{eqnarray}
   \Xi_{\mu\nu\mu'} &=& \frac{1}{2} \frac{\partial}{\partial x^\mu} \square g_{\mu'\nu}
   - \frac{1}{2} \frac{\partial^2}{\partial x^\mu \partial x^{\mu'}}
   \frac{\partial g_{\nu\rho}}{\partial x_\rho} \nonumber \\
   && \hspace{-3em} - \, \frac{1}{2} \Gamma^\sigma_{\mu'\mu} \bigg(
   \frac{1}{\tilde{g}} \square g_{\sigma\nu}
   + \frac{\partial}{\partial x^\nu } \frac{\partial g_{\sigma\rho}}{\partial x_\rho}
   - \frac{\partial}{\partial x^\sigma } \frac{\partial g_{\nu\rho}}{\partial x_\rho}
   \bigg) \nonumber \\
   && \hspace{-3em} + \, \frac{\eta^{\rho\rho'}}{2} \Gamma^\sigma_{\rho\nu} \bigg(
   \frac{\partial^2 g_{\sigma\rho'}}{\partial x^\mu \partial x^{\mu'}}
   - \frac{\partial^2 g_{\mu\rho'}}{\partial x^\sigma \partial x^{\mu'}} \nonumber \\
   && \hspace{0.5em} + \,
   2 \frac{\partial^2 g_{\mu\mu'}}{\partial x^\sigma \partial x^{\rho'}}
   - 2 \frac{\partial^2 g_{\sigma\mu'}}{\partial x^\mu \partial x^{\rho'}}
   - \frac{1}{\tilde{g}} \frac{\partial^2 g_{\sigma\mu}}{\partial x^{\mu'} \partial x^{\rho'}}
   \bigg) \nonumber \\
   && \hspace{-3em} + \, \frac{\eta^{\rho\rho'}}{\tilde{g}} \bigg[ 
   \Gamma^\alpha_{\mu'\mu} \bigg( 
   2 \tilde{\Gamma}_{\alpha\rho'\beta} + \tilde{\Gamma}_{\beta\rho'\alpha}\bigg)
   - \Gamma^\alpha_{\rho'\mu} \tilde{\Gamma}_{\alpha\mu'\beta}
   \bigg] \Gamma^\beta_{\rho\nu} \nonumber \\
   && \hspace{8em} - \; \boxed{\mu \leftrightarrow \nu} = 0 .
\label{compactgeq}
\end{eqnarray}
In view of the antisymmetry of $\Xi_{\mu\nu\mu'}$ in $\mu$ and $\nu$ implied by the last line of the above equation, we can assume $\mu < \nu$ so that Eq.~(\ref{compactgeq}) provides a total of $24$ equations for the ten components of the symmetric matrix $g_{\mu\nu}$. If we wish to determine the time evolution of the metric from third-order differential equations, we need $30$ initial conditions for the ten matrix elements $g_{\mu\nu}$ and their first and second time derivatives as well as expressions for the third time derivatives for all components $g_{\mu\nu}$.

Closer inspection of the third-order terms in Eq.~(\ref{compactgeq}) (see App.~\ref{Appclassfieldeqs}) reveals that the six equations $\Xi_{0 m n} = 0$ for $m \le n$ provide the derivatives $\partial^3 g_{mn}/\partial t^3$, but that the remaining equations do not contain any information about $\partial^3 g_{0\mu}/\partial t^3$. Therefore, the remaining $18$ equations constitute constraints for the initial conditions. As we have realized in Sec.~\ref{seccoco}, the missing  evolution equations for $g_{0\mu}$ are given by the coordinate conditions (\ref{coco3mat}). Equation (\ref{coco3mati}) represents a further constraint.

Although we have identified ten evolution equations for the ten components of the metric $g_{\mu\nu}$ and a total of $19$ primary constraints, it is not clear how may initial conditions remain independent. By differentiating low-order constraints with respect to time and by requiring the dynamic invariance of the constraints we obtain secondary and higher constraints that reduce the number of degrees of freedom considerably below $30-19=11$. However, it is difficult to estimate the exact number of remaining degrees of freedom in composite gravity (or even to guarantee that any degrees of freedom are left). As the handling of constraints is most transparent in the Hamiltonian approach, we next discuss the natural Hamiltonian setting for composite theories \cite{hco237,hco240}.

\section{Hamiltonian formulation}\label{secHamiltonian}
In this section, we tell the story of composite gravity a third time, now from the perspective of the canonical Hamiltonian approach. This approach has three major advantages: (i) it clarifies the structure of the constraints and allows us to count the number of degrees of freedom of the theory, (ii) it facilitates the discussion and taming of Ostrogradsky instabilities \cite{Ostrogradsky1850,Woodard15} in higher derivative theories, and (iii) it provides the key to quantization of the theory. Note that the Hamiltonian approach also guarantees energy conservation and it provides the natural starting point for a generalization to dissipative systems. In particular, this approach allows us to formulate quantum master equations \cite{BreuerPetru,Weiss,hco199,hco221} and to make composite gravity accessible to the robust framework of dissipative quantum field theory \cite{hcoqft,hco243}.

A successful Hamiltonian formulation of a theory benefits from the proper choice of canonical variables. Factorization of the metric is a celebrated strategy for this purpose \cite{Ashtekar86,Ashtekar87}. Whereas one usually avoids too many degrees of freedom by choosing a small symmetry group associated with the decomposition of the metric, composite gravity is built on the six-parameter Lorentz group and a further selection principle for the physically relevant solutions, thus keeping the number of physical degrees of freedom appropriately small for a theory of gravity.

So far, the canonical Hamiltonian formulation of composite gravity in the presence of matter has been elaborated only in the weak-field approximation \cite{hco240}. We here generalize the Hamiltonian approach to the fully nonlinear theory presented in Sec.~\ref{secoverview} and expressed in terms of third-order differential equations for the metric in Sec.~\ref{secmetricform}.

\subsection{Canonical variables}
The fact that we have imposed the third-order coordinate conditions (\ref{coco3mat}) is an obstacle for the Hamiltonian approach, as that approach deals with systems of first-order differential equations. We need to reduce the coordinate condition to first-order equations, at the expense of introducing additional fields. We suggest the following integrated version of the conditions (\ref{coco3mat}),
\begin{equation}\label{coco3mat2i}
   \eta^{\rho\sigma} \, \tilde{\Gamma}_{\mu\rho\sigma} =
   \frac{\partial g_{\mu\rho}}{\partial x_\rho}
   - \frac{1}{2} \frac{\partial {g_\rho}^\rho}{\partial x^\mu} =
   \frac{\partial \phi}{\partial x^\mu} ,
\end{equation}
where the dimensionless scalar field $\phi$ satisfies the second-order differential equation
\begin{equation}\label{potfromT}
   \square \phi = \frac{4 \pi G}{c^4} \, {T_\rho}^\rho
   - \frac{\Lambda}{2} \, {g_\rho}^\rho .
\end{equation}
Equations (\ref{coco3mat2i}) and (\ref{potfromT}) do not only imply the third-order coordinate conditions (\ref{coco3mat}), but also the second-order consistency condition (\ref{coco3mati}). Moreover, Eq.~(\ref{coco3mat2i}) implies the integrability condition
\begin{equation}\label{stronggcond}
   \frac{\partial}{\partial x_\rho} \left( \frac{\partial g_{\nu\rho}}{\partial x^\mu}
   - \frac{\partial g_{\mu\rho}}{\partial x^\nu} \right) = 
   \frac{\partial}{\partial x_\rho} \left( \tilde{\Gamma}_{\nu\mu\rho}
   - \tilde{\Gamma}_{\mu\nu\rho} \right) = 0 .
\end{equation}
For dimensional reasons, and as we do not want conjugate momenta to occur in the coordinate conditions (\ref{coco3mat2i}) through $\partial\phi/\partial t$, the proper Hamiltonian formulation of the above equations requires another scalar configurational field $\bar{\phi}$ ($=\partial\phi/c\partial t$).

An overview of the complete list of variables for the canonical Hamiltonian formulation of composite gravity is given in Table~\ref{tabcanonvariables}. Following the general ideas for the natural Hamiltonian formulation of composite theories \cite{hco237}, the configurational variables are given by the gauge vector fields $A_{a \mu}$ of the Yang-Mills theory based on the Lorentz group and the tetrad variables ${b^\kappa}_\mu$ used in the composition rule (\ref{compositionrule}) for constructing the gauge vector fields. The respective conjugate momenta are denoted by $E^{a \mu}$ and ${p_\kappa}^\mu$, where it is convenient to express conjugate momenta in units of $\hbar$. Moreover, we have introduced the two scalar fields $\phi$, $\bar{\phi}$ and their conjugate momenta $\psi$, $\bar{\psi}$. These scalar fields are essential for the coupling of the gravitational field to matter according to Eqs.~(\ref{coco3mat2i}), (\ref{potfromT}).

\begin{table}
\begin{tabular}{l|l l}
    Yang-Mills variables \hspace{1em}  & $\quad A_{a \mu}$ & $\hbar E^{a \mu}$ \\
	tetrad variables & $\quad {b^\kappa}_\mu$ & $\hbar {p_\kappa}^\mu$ \\
    scalar variables & $\quad \phi$, $\bar{\phi}\qquad$ & $\hbar \psi$, $\hbar \bar{\psi}$ \\
\end{tabular}
\caption{Configurational variables and their conjugate momenta for the canonical Hamiltonian formulation of composite gravity.}
\label{tabcanonvariables}
\end{table}

The canonical Hamiltonian evolution equations for the tetrad variables and their conjugate momenta are given by
\begin{equation}\label{Hamtetradeqs}
   \frac{\partial {b^\kappa}_\mu}{\partial t} =
   \frac{1}{\hbar} \frac{\delta H}{\delta {p_\kappa}^\mu} ,
   \qquad
   \frac{\partial {p_\kappa}^\mu}{\partial t} =
   - \frac{1}{\hbar} \frac{\delta H}{\delta {b^\kappa}_\mu} ,
\end{equation}
where $H$ is the total Hamiltonian. The evolution equations for the Yang-Mills fields are obtained from the following equations (note the unusual sign conventions in addition to the factors of $1/\hbar$),
\begin{equation}\label{HamYMeqs}
   \frac{\partial A_{a \mu}}{\partial t} = - \frac{1}{\hbar} \frac{\delta H}{\delta E^{a \mu}} ,
   \qquad
   \frac{\partial E^{a \mu}}{\partial t} = \frac{1}{\hbar} \frac{\delta H}{\delta A_{a \mu}} .
\end{equation}
Finally, the scalar fields are governed by the canonical evolution equations
\begin{equation}\label{Hamscaleqs}
   \frac{\partial \phi}{\partial t} = \frac{1}{\hbar} \frac{\delta H}{\delta \psi} ,
   \qquad
   \frac{\partial \psi}{\partial t} = - \frac{1}{\hbar} \frac{\delta H}{\delta \phi} ,
\end{equation}
and analogously
\begin{equation}\label{Hamscaleqsb}
   \frac{\partial \bar{\phi}}{\partial t} = \frac{1}{\hbar} \frac{\delta H}{\delta \bar{\psi}} ,
   \qquad
   \frac{\partial \bar{\psi}}{\partial t} = - \frac{1}{\hbar} \frac{\delta H}{\delta \bar{\phi}} .
\end{equation}
The concrete form of all these evolution equations can be written down as soon as we have specified the total Hamiltonian. In particular, we need to reproduce all the evolution equations contained in the composition rule (\ref{compositionrule}), in the coordinate conditions (\ref{coco3mat2i}), (\ref{potfromT}), and in the Yang-Mills field equations (\ref{YMfieldeqs}).

\subsection{Hamiltonian}
The full Hamiltonian $H$ consists of the matter contribution $H_{\rm m}$ given in Eq.~(\ref{particleH}) and a field contribution $H_{\rm f}$. In the spirit of composite theories \cite{hco237}, we write
\begin{eqnarray}
   H_{\rm f} &=& \hbar c \int \bigg[ \frac{1}{2} E^{a \mu} E_{a \mu}
   + \frac{1}{4} F_{a mn} F^{a mn}
   - E^{a 0} \frac{\partial A_{a n}}{\partial x_n} \nonumber \\
   &-& E^{a n} \left( \frac{\partial A_{a 0}}{\partial x^n}
   + f_a^{bc} A_{b n} A_{c 0} \right)
   + X_{\mu\nu} \, \bar{b}^{\mu\kappa} {p_\kappa}^\nu \nonumber \\
   &+& \psi \bar{\phi} + \bar{\psi} \bigg( \frac{\partial^2 \phi}{\partial x_n \partial x^n} 
   - \frac{4 \pi G}{c^4} \, {T_\rho}^\rho + \frac{\Lambda}{2} \, {g_\rho}^\rho
   \bigg) \bigg] \, d^3 x . \qquad \quad 
\label{Hfield}
\end{eqnarray}
The terms containing $A_{a \mu}$, $E^{a \mu}$ and $F_{a mn}$ represent the standard Hamiltonian of Yang-Mills theory with a gauge-breaking term corresponding to the Feynman gauge (see, e.g., Sec.~2.2 of \cite{BassettoNardelliSoldati}, Chap.~15 of \cite{WeinbergQFT2}, or \cite{hco229}). The coupling between Yang-Mills and tetrad variables is implemented by the term containing $X_{\mu\nu}$. The remaining terms in Eq.~(\ref{Hfield}) are associated with the scalar variables and, in particular, the coupling to matter through the trace of the energy-momentum tensor. The quantities $X_{\mu\nu}$, which we assume to be independent of the variables ${p_\kappa}^\mu$ (they actually turn out to depend only on the configurational variables $A_{a \mu}$, ${b^\kappa}_\mu$, $\phi$ and $\bar{\phi}$), remain to be determined from the evolution equations for the tetrad variables resulting from Eqs.~(\ref{Hamtetradeqs}) and (\ref{Hfield}),
\begin{equation}\label{bevol}
   \frac{1}{c} \frac{\partial {b^\kappa}_\mu}{\partial t} = \bar{b}^{\nu\kappa} X_{\nu\mu} ,
   \quad \mbox{or} \qquad
   X_{\mu\nu} = b_{\kappa\mu} \frac{\partial {b^\kappa}_\nu}{\partial x^0} .
\end{equation}
According to the product rule, the quantities $X_{\mu\nu}$ have the convenient property
\begin{equation}\label{gevol}
   \frac{\partial g_{\mu\nu}}{\partial x^0} = X_{\mu\nu} + X_{\nu\mu} .
\end{equation}
Only the antisymmetric part of $X_{\mu\nu}$ depends on the particular choice of the decomposition (\ref{localinertialdecomp}) of the metric, that is, on the gauge.

The Hamiltonian (\ref{Hfield}) has the serious problem that it is not bounded from below. This may lead to so-called Ostrogradsky instabilities, which are well-known for higher derivative theories \cite{Ostrogradsky1850,Woodard15}. The lack of a lower bound is most obvious from the occurrence of terms that are linear in the conjugate momenta, but there are also possible issues with the signs of quadratic terms. Avoiding such instabilities is an important topic in higher derivative theories, in particular, in alternative theories of gravity \cite{Chenetal13,RaidalVeermae17,Stelle77,Stelle78,Krasnikov87,GrosseKnetter94,Beckeretal17,Salvio19}. We here avoid instabilities by imposing constraints, which is a promising strategy for composite higher derivative theories \cite{hco235,hco237}. A systematic discussion of these constraints is postponed to Sec.~\ref{secHamconstraintsstab}.

To complete the explicit form of the Hamiltonian $H_{\rm f}$, we determine the quantities $X_{\mu\nu}$ from Eq.~(\ref{bevol}). By using the definition (\ref{Atildef}) to evaluate
$2 \tilde{A}_{mn0} + \tilde{A}_{0mn} + \tilde{A}_{0nm}$,
we find the explicit expression
\begin{eqnarray}\label{Xeqsmn}
   X_{mn} &=& 
   \frac{\partial g_{0m}}{\partial x^n}
   + {b^\kappa}_m {b^\lambda}_n \, A_{(\kappa\lambda) 0} \\
   && \hspace{-3em} + \, \frac{1}{2} {b^\kappa}_0 \bigg( {b^\lambda}_m \, A_{(\kappa\lambda) n}
   + {b^\lambda}_n \, A_{(\kappa\lambda) m}
   - \frac{\partial b_{\kappa m}}{\partial x^n}
   - \frac{\partial b_{\kappa n}}{\partial x^m} \bigg) . \nonumber
\end{eqnarray}
From Eq.~(\ref{Atildef}) for $\mu = \rho = 0$, $\nu \neq 0$, we further obtain the compact result
\begin{equation}\label{Xeqs0n}
   X_{0n} = \frac{1}{2} \frac{\partial g_{00}}{\partial x^n}
   + {b^\kappa}_0 {b^\lambda}_n \, A_{(\kappa\lambda) 0} .
\end{equation}
We still need an expression for $X_{\mu 0}$, which cannot be extracted from the composition rule. The missing four conditions are provided by the coordinate conditions (\ref{coco3mat2i}), which lead to
\begin{equation}\label{Xeqs0ns}
   X_{n0} = \frac{\partial g_{mn}}{\partial x_m}
   - \frac{1}{2} \, \frac{\partial g_{mm}}{\partial x_n}
   - {b^\kappa}_0 {b^\lambda}_n \, A_{(\kappa\lambda) 0}
   - \frac{\partial\phi}{\partial x^n} ,
\end{equation}
and\begin{equation}\label{Xeqs00}
   X_{00} = {b^\kappa}_0 \bigg( \frac{\partial b_{\kappa n}}{\partial x_n}
   - {b^\lambda}_n \, A_{(\kappa\lambda) n}  \bigg) - \bar{\phi} ,
\end{equation}
where we have introduced the additional scalar configurational variable $\bar{\phi} = \partial\phi/\partial x^0$, which is consistent with the Hamiltonian dynamics. At this point, the definition of the Hamiltonian (\ref{Hfield}) is completed. As anticipated, the quantities $X_{\mu\nu}$ in Eqs.~(\ref{Xeqsmn})--(\ref{Xeqs00}) depend only on the configurational fields $A_{a \mu}$, ${b^\kappa}_\mu$, $\phi$ and $\bar{\phi}$---including spatial derivatives of ${b^\kappa}_\mu$ and $\phi$. The fact that the structure of the expressions for $X_{\mu 0}$ matches the structure of $X_{mn}$ and $X_{0n}$ so nicely is a further argument in favor of the coordinate conditions (\ref{coco3mat2i}). The additional occurrence of the scalar fields provides the coupling of the gravitational field to the energy-momentum tensor of matter.

We might now be tempted to take the total Hamiltonian $H$ simply as the sum $H_{\rm f} + H_{\rm m}$. However, one should notice that, on the one hand, $H_{\rm m}$ corresponds to the macroscopic energy of a mass point $m$, which could typically represent a planet, a star, or even an entire galaxy. On the other hand, $H_{\rm f}$ corresponds to the energy of quanta of the gravitational field. We hence assume a total Hamiltonian of the form
\begin{equation}\label{Htot}
   H = H_{\rm f} + \Lambda_{\rm E} \, H_{\rm m} ,
\end{equation}
where $\Lambda_{\rm E}$ is an extremely small dimensionless constant. In \cite{hco240}, the parameter $\Lambda_{\rm E}$ has been estimated to be of order $10^{-124}$. Note that this value is of the order of the cosmological constant when distances are measured in units of the Planck length. Quite remarkably, the insistence on a Hamiltonian description of the combined system of gravitational field and matter necessitates an exceedingly small but nonzero number $\Lambda_{\rm E}$.

\subsection{Evolution equations}\label{secHamevoleqs}
So far, we have considered only the evolution equations (\ref{bevol}) for the tetrad variables. These equations were used to construct the quantities $X_{\mu\nu}$ in the Hamiltonian (\ref{Hfield}) such that the twelve evolution equations contained in the composition rule (\ref{compositionrule}) and the four coordinate conditions (\ref{coco3mat2i}) are reproduced. We next consider all the remaining evolution equations for the fields listed in Table~\ref{tabcanonvariables}.

Equations (\ref{Hamscaleqs}) and (\ref{Hamscaleqsb}) lead to the following evolution equations for the scalar fields,
\begin{equation}\label{phievol}
   \frac{\partial \phi}{\partial x^0} = \bar{\phi} ,
\end{equation}
\begin{equation}\label{phibevol}
   \frac{\partial \bar{\phi}}{\partial x^0} =
   \frac{\partial^2 \phi}{\partial x_n \partial x^n} 
    - \frac{4 \pi G}{c^4} \, {T_\rho}^\rho + \frac{\Lambda}{2} \, {g_\rho}^\rho ,
\end{equation}
and for their conjugate momenta we obtain
\begin{equation}\label{pibevol}
   \frac{\partial \bar{\psi}}{\partial x^0} = - \psi + \bar{b}^{0\kappa} \, {p_\kappa}^0 ,
\end{equation}
\begin{equation}\label{pievol}
   \frac{\partial \psi}{\partial x^0} =
   - \frac{\partial^2 \bar{\psi}}{\partial x_n \partial x^n} 
   - \frac{\partial}{\partial x^n} ( \bar{b}^{n\kappa} \, {p_\kappa}^0 ) .
\end{equation}
Equations (\ref{phievol}) and (\ref{phibevol}) imply the second-order partial differential equation (\ref{potfromT}) for $\phi$, whereas Eqs.~(\ref{pibevol}) and (\ref{pievol}) can be combined into the second-order field equation
\begin{equation}\label{pibevol2}
   \square \bar{\psi} + \frac{\partial}{\partial x^\mu} ( \bar{b}^{\mu\kappa} \, {p_\kappa}^0 ) = 0 .
\end{equation}

For the evolution of the gauge vector fields, Eq.~(\ref{HamYMeqs}) implies
\begin{equation}\label{Aevola0}
   \frac{\partial A_{a 0}}{\partial x^0} = - E_{a 0} + \frac{\partial A_{a n}}{\partial x_n} ,
\end{equation}
and
\begin{equation}\label{Aevolaj}
   \frac{\partial A_{a m}}{\partial x^0} = - E_{a m} + \frac{\partial A_{a 0}}{\partial x^m}
   + f_a^{bc} A_{b m} A_{c 0} .
\end{equation}
These equations provide the relationship between the conjugate momenta $E_{a \mu}$ and the time derivatives of the gauge vector fields. Equation (\ref{Aevola0}) implies $E_{a 0} = \partial A_{a \mu}/\partial x_\mu$, so that the Lorenz gauge condition (\ref{Lorenzgauge}) can be rewritten as $E_{a 0} = 0$. Equation (\ref{Aevolaj}) implies that the spatial components of the conjugate momenta are given by the mixed space-time components of the field tensor, $E_{a j} = F_{a j0} = - F_{a 0j}$, and hence the analogues of the electric fields in the theory of electromagnetism.

The components $E_{a j}$ inherit the gauge-transformation behavior (\ref{gaugeF}) of the field tensor. The transformation behavior of  $E_{a 0}$ can only be specified if we lift the Lorenz gauge condition and provide evolution equations for the gauge transformations, which thus become part of the dynamic system. Unlike the $A_{a \mu}$, their conjugate fields $E_{a \mu}$ are not the components of a four-vector. Likewise, we expect the fields ${p_\kappa}^\mu$ neither to be four-vectors nor to possess obvious gauge-transformation behavior.

The evolution of the conjugate momenta $E_{a \mu}$ is given by Eq.~(\ref{HamYMeqs}),
\begin{equation}\label{Eevola0}
   \frac{\partial E^a_0}{\partial x^0} = - \frac{\partial E^a_n}{\partial x_n}
   - f_c^{ab} A_{b n} E^{c n} - J^a_0 ,
\end{equation}
and
\begin{eqnarray}
   \frac{\partial E^a_m}{\partial x^0} &=& - \frac{\partial E^a_0}{\partial x^m}
   - f_c^{ab} A_{b 0} E^c_m 
   - \frac{\partial^2 A^a_m}{\partial x^n \partial x_n}
   + \frac{\partial^2 A^a_n}{\partial x^m \partial x_n} \nonumber \\
   &+& f^{abc} \left( A_{b n} \frac{\partial A_{c n}}{\partial x_m}
   + A_{b m} \frac{\partial A_{c n}}{\partial x_n}
   -2 A_{b n} \frac{\partial A_{c m}}{\partial x_n} \right) \nonumber \\
   &-& f_s^{ab} f^{scd} A_{b n} A_{c n} A_{d m} - J^a_m ,
\label{Eevolaj}
\end{eqnarray}
with the following definitions of the currents $J^a_\mu$,
\begin{equation}\label{Ja0expr}
   J^{(\kappa\lambda)}_0 = p^{\kappa\mu} \, {b^\lambda}_\mu - {b^\kappa}_\mu \, p^{\lambda\mu} ,
\end{equation}
\begin{eqnarray}
   J^{(\kappa\lambda)}_m &=& \frac{1}{2} 
   \big( {b^\kappa}_n {b^\lambda}_0 - {b^\kappa}_0 {b^\lambda}_n \big)  \big( \mbox{$\bar{b}^n$}_{\kappa'} {p^{\kappa'}}_m
   + \bar{b}_{m \kappa'} p^{\kappa'n} \big) \nonumber\\
   &+& \big( {b^\kappa}_m {b^\lambda}_0 - {b^\kappa}_0 {b^\lambda}_m \big) \,
   \bar{b}_{0 \kappa'} \, p^{\kappa'0} .
\label{Jajexpr}
\end{eqnarray}
In general, these currents $J^a_\mu$ must be Lorentz four-vectors and must possess the proper gauge-transformation behavior implied by Eqs.~(\ref{YMfieldeqs}) and (\ref{gaugefieldeqs}).

To simplify the further discussion of the relationship between the currents $J^a_\mu$ and the conjugate momenta ${p_\kappa}^\mu$, we introduce the auxiliary quantities
\begin{equation}\label{Jaux}
   \hat{J}^{\mu\nu}_\rho =
   \mbox{$\bar{b}^\mu$}_{\kappa} \mbox{$\bar{b}^\nu$}_{\lambda} \,
    J^{(\kappa\lambda)}_\rho ,
\end{equation}
and
\begin{equation}\label{paux}
   \hat{p}^{\mu\nu} = \bar{b}^{\mu\kappa} \, {p_\kappa}^\nu .
\end{equation}
For every pair $\mu \neq \nu$, $\hat{J}^{\mu\nu}_\rho$ is a gauge-invariant four-vector field. A local conservation law for $\hat{J}^{\mu\nu}_\rho$ is given in Eq.~(\ref{Jhatconservation}) or (\ref{Jhatconservationx}). The above definitions allow us to rewrite Eqs.~(\ref{Ja0expr}) and (\ref{Jajexpr}) in the more transparent form
\begin{equation}\label{Ja0exprhat}
   \hat{J}^{\mu\nu}_0 = \hat{p}^{\mu\nu} - \hat{p}^{\nu\mu} ,
\end{equation}
and
\begin{equation}\label{Jajexprhat}
   \hat{J}^{0n}_m = - \frac{1}{2} \big( \hat{p}^{mn} + \hat{p}^{nm} \big)
   + \delta_{mn} \, \hat{p}^{00} ,
\end{equation}
where in the latter equation only the nonzero components are displayed (except for those given by the obvious antisymmetry $\hat{J}^{n0}_m=-\hat{J}^{0n}_m$). As identically vanishing spatial components must be accompanied by vanishing temporal components, we obtain the symmetry requirement
\begin{equation}\label{phatsym}
   \hat{p}^{mn} = \hat{p}^{nm} .
\end{equation}
Conversely, $\hat{J}^{0n}_0 = \hat{p}^{0n} - \hat{p}^{n0}$ cannot be identically zero.

We finally consider the evolution equations for the conjugate momenta ${p_\kappa}^\mu$, which we write as equations for $\hat{p}^{\mu\nu}$. From Eq.~(\ref{Hamtetradeqs}), we obtain
\begin{equation}\label{Hamphat}
   \frac{\partial \hat{p}^{\mu\nu}}{\partial x^0} =
   - \bar{g}^{\mu\rho} X_{\rho\sigma} \hat{p}^{\sigma\nu}
   - \frac{1}{\hbar c} \, \bar{b}^{\mu\kappa} \frac{\delta H}{\delta {b^\kappa}_\nu} .
\end{equation}
If we introduce the symmetric tensors associated with the energy-momentum tensor defined in Eq.~(\ref{particleT}),
\begin{equation}\label{Thatdef}
   \hat{T}^{\mu\nu} = \bar{g}^{\mu\rho} \, {T_\rho}^\nu ,
\end{equation}
the remaining evolution equations can be written as
\begin{eqnarray}
   \frac{\partial \hat{p}^{\mu 0}}{\partial x^0} &=&
   \frac{\partial \hat{p}^{\mu n}}{\partial x^n}
   + \Gamma^\mu_{mn} \, \hat{p}^{mn} - \Gamma^\mu_{nn} \, \hat{p}^{00}
   \nonumber\\
   &+& \bar{g}^{\mu\nu} ( X_{\nu n} - \tilde{A}_{\nu n0} ) ( \hat{p}^{0n} - \hat{p}^{n 0} )
   \nonumber\\
   &-& \frac{2 \pi G}{c^4} \bar{\psi} \bar{g}^{\mu0} \, \hat{T}^{00}
   - \bar{\psi} \Lambda \, \eta^{\mu 0}
   + \frac{\Lambda_{\rm E}}{\hbar c} \, \hat{T}^{\mu 0} ,
\label{pmu0evol}
\end{eqnarray}
\begin{eqnarray}
   \frac{\partial \hat{p}^{0n}}{\partial x^0} &=&
   \frac{\partial \hat{p}^{00}}{\partial x^n}
   - \Gamma^0_{l0} \, \hat{p}^{ln} + \Gamma^0_{n0} \, \hat{p}^{00}
   \nonumber\\
   &-& \bar{g}^{0\nu} ( X_{\nu 0} - \tilde{A}_{\nu 00} ) ( \hat{p}^{0n} - \hat{p}^{n 0} )
   \nonumber\\
   &-& \frac{2 \pi G}{c^4} \bar{\psi} \bar{g}^{0n} \, \hat{T}^{00}
   + \frac{\Lambda_{\rm E}}{\hbar c} \, \hat{T}^{0n} ,
\label{p0nevol}
\end{eqnarray}
and
\begin{eqnarray}
   \frac{\partial \hat{p}^{m n}}{\partial x^0} &=&
   \frac{\partial \hat{p}^{m0}}{\partial x^n} + \frac{\partial \hat{p}^{n0}}{\partial x^m}
   - \Gamma^m_{l0} \, \hat{p}^{ln} + \Gamma^m_{n0} \, \hat{p}^{00}
   \nonumber\\
   &-& \bar{g}^{m \nu} ( X_{\nu 0} - \tilde{A}_{\nu 00} ) ( \hat{p}^{0n} - \hat{p}^{n 0} )
   \nonumber\\
   &-& \frac{2 \pi G}{c^4} \bar{\psi} \left( \bar{g}^{0m} \hat{T}^{0n}
   + \bar{g}^{0n} \hat{T}^{0m} - \bar{g}^{00} \hat{T}^{mn} \right)
   \nonumber \\
   &-& \left( \bar{\psi} \Lambda + \frac{\partial \hat{p}^{l0}}{\partial x^l} \right) \, \eta^{m n}
   + \frac{\Lambda_{\rm E}}{\hbar c} \, \hat{T}^{mn} .
\label{pmnevol}
\end{eqnarray}
These evolution equations for the fluxes still need to be better appreciated. In Sections~\ref{secoverview} and \ref{secmetricform}, we considered only vanishing fluxes.

\subsection{Constraints for stability}\label{secHamconstraintsstab}
If we impose the constraint $\bar{\psi}=0$, we can kill two birds with one stone: we eliminate the linear dependence of the Hamiltonian $H_{\rm f}$ in Eq.~(\ref{Hfield}) on $\bar{\psi}$, which is a source of Ostrogradsky instability, and we make sure that $H_{\rm f}$  does not destroy geodesic particle motion obtained from $H_{\rm m}$ in Sec.~\ref{secparticlemotion} through its dependence on ${T_\rho}^\rho$. If $\bar{\psi}=0$ is adopted as a fundamental constraint, Eqs.~(\ref{pibevol}) and (\ref{pibevol2}) leave us with two options:
\begin{equation}\label{fundconstrs1}
   \bar{\psi}=0 , \qquad \psi = \hat{p}^{00} , \qquad
   \frac{\partial \hat{p}^{\mu 0}}{\partial x^\mu} = 0 ,
\end{equation}
as the weaker set of constraints, or
\begin{equation}\label{fundconstrs2}
   \bar{\psi} = \psi = \hat{p}^{\mu 0} = 0 ,
\end{equation}
as the stronger set of constraints eliminating even more terms potentially leading to instabilities from the Hamiltonian (\ref{Hfield}).  Moreover, the stronger constraints (\ref{fundconstrs2}), which we from now on adopt as primary constraints, supplement the Lorenz gauge $E_{a 0} = 0$ by the analogous constraint $p_{\kappa 0} = 0$ and lead to a nicely compact relation between the non-vanishing currents and the conjugate momenta,
\begin{equation}\label{Jprel}
   \hat{J}^{0n \mu} = - \hat{p}^{\mu n} .
\end{equation}

For the strong constraints (\ref{fundconstrs2}), the evolution equations (\ref{p0nevol}) and (\ref{pmnevol}) can, with the help of the coordinate conditions (\ref{coco3mat2i}), be combined into the compact form
\begin{equation}\label{pmunevolx}
   \frac{\partial\hat{J}^{0n \mu}}{\partial x^0} + \Gamma^\mu_{0 \nu} \, \hat{J}^{0n \nu} =
   - \frac{\Lambda_{\rm E}}{\hbar c} \, \hat{T}^{\mu n} ,
\end{equation}
whereas Eq.~(\ref{pmu0evol}) becomes the constraint
\begin{equation}\label{pmu0evolx}
   \frac{\partial \hat{J}^{0n \mu}}{\partial x^n} + \Gamma^\mu_{n \nu} \,\hat{J}^{0n \nu} =
   \frac{\Lambda_{\rm E}}{\hbar c} \, \hat{T}^{\mu 0} .
\end{equation}
The time invariance of this constraint can be expressed as
\begin{equation}\label{pmu0evolx2}
   {R^\mu}_{\nu 0n} \, \hat{J}^{0n \nu} = - \frac{\Lambda_{\rm E}}{\hbar c}
   \left( \frac{\partial \hat{T}^{\mu\nu}}{\partial x^\nu}
   + \Gamma^\mu_{\nu \rho} \, \hat{T}^{\rho\nu} \right) = 0 ,
\end{equation}
where the expression in parentheses vanishes because of the energy-momentum conservation (\ref{particleTdivg}). Dynamic invariance of the symmetry (\ref{phatsym}) requires the constraint
\begin{equation}\label{phatsym2}
   \Gamma^m_{0 \nu} \, \hat{J}^{0n \nu} = \Gamma^n_{0 \nu} \, \hat{J}^{0m \nu} .
\end{equation}

Equations (\ref{pmunevolx}) and (\ref{pmu0evolx}) show that the idea of vanishing external fluxes, which we have used in Sections~\ref{secoverview} and \ref{secmetricform}, is not strictly possible in the presence of matter. However, as $\Lambda_{\rm E}$ is an extremely small parameter, non-vanishing fluxes can matter only on cosmological scales. As we are primarily interested in the small scales relevant to quantum effects, we still assume that all fluxes $\hat{J}^{\mu\nu}_\rho$ vanish. More formally, we are interested in the limit $\Lambda_{\rm E} \rightarrow 0$. For the linearized theory \cite{hco240}, it has been shown that, if external fluxes are admitted in the Yang-Mills theory, also the equations for the tetrad variables need to be modified for reasons of consistency. The nonlinear composite theory of gravity on cosmological length and mass scales presumably requires additional insights and still remains to be developed. It would be over-optimistic to expect that a theory of gravity modified for improved behavior at small length scales would automatically improve the behavior at large length scales.

\subsection{Gauge constraints}\label{secHamconstraintsgauge}
Just to offer a flavor of the kind and number of the gauge constraints, we here consider the linearized version of composite gravity. The gauge constraints for the nonlinear theory should be treated within the less transparent BRST formalism \cite{BecchiRouetStora76,Tyutin75,Nemeschanskyetal86,hco229}. We here do not want to go into further details of this formalism, which is well-established but requires additional variables (ghost fields).

According to Eq.~(\ref{Aevola0}), the Lorenz gauge (\ref{Lorenzgauge}) can be expressed in the form of the simple constraint
\begin{equation}\label{gaugeconstrEa01}
    E^a_0 = 0 .
\end{equation}
This gauge condition is known to be incomplete. In order to fix the gauge completely, one often imposes the additional constraint
\begin{equation}\label{gaugeconstrAa01}
    A_{a0} = 0 .
\end{equation}
According to Eqs.~(\ref{Aevola0}) and (\ref{Eevola0}), the secondary constraints obtained from the time independence of these gauge constraints for the linearized theory are given by
\begin{equation}\label{gaugeconstrEa02}
    \frac{\partial E^a_n}{\partial x_n} = 0 ,
\end{equation}
and
\begin{equation}\label{gaugeconstrAa02}
    \frac{\partial A_{a n}}{\partial x_n} = 0 ,
\end{equation}
respectively, where the latter is known as Coulomb gauge. These secondary constraints are invariant under the time-evolution equations (\ref{Aevolaj}) and (\ref{Eevolaj}).

The most important feature for us is that the $24$ constraints  (\ref{gaugeconstrEa01})--(\ref{gaugeconstrAa02})  reduce the number of degrees of freedom of the Yang-Mills theory by one half by eliminating the temporal and longitudinal degrees of freedom. This count holds also for the nonlinear theory.

\subsection{Constraints from composition rule}\label{secHamconstraintscomp}
We next turn to another source of primary constraints, namely the composition rule (\ref{compositionrule}) or (\ref{Atildef}). For the discussion of these constraints it is useful to introduce the gauge-dependent quantities
\begin{equation}\label{Omders}
   \Omega_{\mu\nu/\rho} = {b^\kappa}_\mu \, \frac{\partial b_{\kappa\nu}}{\partial x^\rho}
      - \frac{\partial{b^\kappa}_\mu}{\partial x^\rho} \, b_{\kappa\nu} ,
\end{equation}
which are antisymmetric in $\mu$ and $\nu$.
From the composition rule (\ref{Atildef}) for $\tilde{g}=1$ we can extract the $12=9+3$ primary constraints (cf.\ Eqs.~(23) and (24) of \cite{hco240})
\begin{equation}\label{YMqlinconstraints1a}
   \tilde{A}_{mn j} =
   \frac{1}{2} \left( \frac{\partial g_{jn}}{\partial x^m}
   - \frac{\partial g_{jm}}{\partial x^n} + \Omega_{mn/j} \right) ,
\end{equation}
and
\begin{equation}\label{YMqlinconstraints1b}
   \tilde{A}_{m0n} - \tilde{A}_{n0m} =
   \frac{1}{2} \bigg( \frac{\partial g_{0n}}{\partial x^m}
   - \frac{\partial g_{0m}}{\partial x^n} + \Omega_{0n/m} - \Omega_{0m/n} \bigg) .
\end{equation}
By means of Eq.~(\ref{Atildefx}), these primary constraints can be rewritten in a more compact, manifestly gauge-invariant form in terms of connection variables,
\begin{equation}\label{METqlinconstraints1a}
   \tilde{\Gamma}_{mnj} = \frac{1}{2} \left( \frac{\partial g_{jm}}{\partial x^n}
   + \frac{\partial g_{mn}}{\partial x^j} - \frac{\partial g_{jn}}{\partial x^m} \right) ,
\end{equation}
and
\begin{equation}\label{METqlinconstraints1b}
   \tilde{\Gamma}_{mn0} - \tilde{\Gamma}_{nm0} =
   \frac{\partial g_{0m}}{\partial x^n} - \frac{\partial g_{0n}}{\partial x^m} .
\end{equation}
The time derivative of the latter form of the primary constraints leads to the following secondary constraints (cf.\ Eqs.~(46) and (47) of \cite{hco240}),
\begin{equation}\label{METqlinconstraints2a}
   \tilde{E}_{mn j} 
   - \Gamma^\sigma_{mj} \tilde{\Gamma}_{\sigma 0n}
   + \Gamma^\sigma_{nj} \tilde{\Gamma}_{\sigma 0m} = 
   \frac{\partial \tilde{\Gamma}_{0nj}}{\partial x^m}
   - \frac{\partial \tilde{\Gamma}_{0mj}}{\partial x^n} ,
\end{equation}
and
\begin{equation}\label{METqlinconstraints2b}
   \tilde{E}_{0m n} = \tilde{E}_{0n m} ,
\end{equation}
where the evolution equations (\ref{Gamtilevol}) have been used. According to Eq.~(\ref{Fdefinitiontilz}), these secondary constraints are contained in the symmetry properties of the field tensor observed after Eq.~(\ref{Fdefinitiontilz}), as $\tilde{E}_{mn j} = -\tilde{F}_{mn0j} = -\tilde{F}_{0jmn}$ and $\tilde{F}_{0m0n} = \tilde{F}_{0n0m}$. The remaining symmetries $\tilde{F}_{mnm'n'} = \tilde{F}_{m'n'mn}$ do not occur as secondary constraints because they involve only spatial derivatives of the composition rule.

The definition (\ref{Fdefinition}) of the field tensor actually consists of $18$ evolution equations [given by Eq.~(\ref{Aevolaj}) or Eq.~(\ref{Gamtilevol})] and $18$ constraints [definitions of $F_{am}$ in terms of gauge vector fields and their spatial derivatives]. The missing $6$ evolution equations are contained in the definitions (\ref{Aevola0}) of the fields $E_{a 0}$ or, eventually, in the Lorenz gauge (\ref{Lorenzgauge}).

From the time derivatives of the secondary constraints (\ref{METqlinconstraints2a}) and (\ref{METqlinconstraints2b}), we obtain the tertiary constraints (cf.\ Eqs.~(89) and (90) of \cite{hco240})
\begin{eqnarray}
   \tilde{J}_{mn j} &+& \frac{\partial}{\partial x^m} \left( \tilde{E}_{0j n}
   + \Gamma^\sigma_{jn} \tilde{\Gamma}_{\sigma 00}
   - \Gamma^\sigma_{0n} \tilde{\Gamma}_{\sigma 0j} \right)
   \nonumber\\
   &-& \frac{\partial}{\partial x^n} \left( \tilde{E}_{0j m}
      + \Gamma^\sigma_{jm} \tilde{\Gamma}_{\sigma 00}
      - \Gamma^\sigma_{0m} \tilde{\Gamma}_{\sigma 0j} \right) = \qquad 
   \nonumber\\
   && \hspace{-1.1em} \frac{\partial \tilde{F}_{jr mn}}{\partial x_r}
   - \frac{\partial}{\partial x^0} \big( \Gamma^\sigma_{jm} \tilde{\Gamma}_{\sigma 0 n} 
   -\Gamma^\sigma_{jn} \tilde{\Gamma}_{\sigma 0 m} \big)
   \nonumber\\
   &+& \eta^{\mu\mu'} \Big( \Gamma^\sigma_{\mu m} \tilde{F}_{\sigma n \mu'j} 
   + \Gamma^\sigma_{\mu n} \tilde{F}_{m \sigma \mu' j} \Big) ,
\label{METqlinconstraints3a}
\end{eqnarray}
and
\begin{eqnarray}
   \frac{\partial\tilde{E}_{mn l}}{\partial x_l} - \tilde{J}_{0mn} + \tilde{J}_{0nm} &=&
   \eta^{\mu\nu} \Big( \Gamma^\sigma_{\mu 0} \tilde{F}_{\nu\sigma mn}
   + \Gamma^l_{\mu m} \tilde{E}_{\nu nl} \nonumber\\
   && \hspace{1.8em} + \Gamma^l_{\mu n} \tilde{E}_{m\nu l} \Big) ,
\label{METqlinconstraints3b}
\end{eqnarray}
where Eqs.~(\ref{EJacobi}) and (\ref{FJacobix}) have been used. These tertiary constraints can actually be recognized as the constraints $\Xi_{mnj}=0$ and $\Xi_{0nm}-\Xi_{0mn}=0$ contained in the field equations (note that the tertiary constraints coincide with the constraints (\ref{geqclass2}) and (\ref{geqclass3}) identified in Appendix~\ref{Appclassfieldeqs}). An overview of how all the equations of the previous sections arise in the Hamiltonian approach is given in Table~\ref{tabeqstatus}.

\begin{table}
\begin{tabular*}{\columnwidth}{l l}
Composition rule (\ref{compositionrule}) & $12$ evolution equations for ${b^\kappa}_n$\\
   & $12$ primary constraints\\
\hline
Coordinate conditions (\ref{coco3mat2i}) & $4$ evolution equations for ${b^\kappa}_0$\\
   \qquad\qquad and (\ref{potfromT}) & and expression for $\square\phi$ \\
\hline
Evolution Eqs.~(\ref{Aevola0}) for $A_{a0}$ &
   Definition $E_{a0}=\partial A_{a\mu}/\partial x_\mu$ \\
   & [$E_{a0}=0$ for Lorenz gauge (\ref{Lorenzgauge})] \\
Evolution Eqs.~(\ref{Aevolaj}) for $A_{am}$ \ & Definition  $E_{am}=F_{am0}$ \\
\hline
Evolution Eqs.~(\ref{Eevola0}) & Field equations $\Xi_{\mu\nu\mu'}=0$ \\
   \qquad\qquad and (\ref{Eevolaj}) for $E^a_{\mu'}$ &
   [with fluxes in terms of $\hat{p}^{\mu\nu}$] \\
\hline
Secondary constraints (\ref{METqlinconstraints2a}) & $12$ symmetry conditions \\
   \qquad\qquad and (\ref{METqlinconstraints2b}) &
   $\tilde{F}_{0m \mu\nu} = \tilde{F}_{\mu\nu 0m}$ \\
Tertiary constraints (\ref{METqlinconstraints3a}) & $12$ field equations $\Xi_{mnj}=0$ \\
   \qquad\qquad and (\ref{METqlinconstraints3b}) &
   and $\Xi_{0nm}-\Xi_{0mn}=0$ \\
\end{tabular*}
\caption{Occurrence of various equations in the canonical Hamiltonian approach.}
\label{tabeqstatus}
\end{table}

\subsection{Number of degrees of freedom}\label{secdof}
We are now in a position to count the number of degrees of freedom in the composite theory of gravity. The Hamiltonian system consists of $84$ first-order evolution equations for the variables listed in Table~\ref{tabcanonvariables}. We have identified $18$ constraints for conjugate momenta to provide stability in Sec.~\ref{secHamconstraintsstab}, $24$ gauge constraints in Sec.~\ref{secHamconstraintsgauge}, and $36$ constraints from the composition rule in Sec.~\ref{secHamconstraintscomp}. In total, we are left with only six degrees of freedom in the composite theory of gravity. This number differs from the four or five degrees of freedom counted in Sec.~III.C of \cite{hco240} because we have chosen different coordinate conditions requiring additional scalar fields.

\section{Static isotropic solution}\label{secisosol}
We are interested in the static isotropic solution for composite gravity because (i) it allows us to verify consistency with the high-precision predictions of general relativity and (ii) it determines the basic features of black holes. We first compile the basic equations characterizing static isotropic solutions and then solve them by series expansions and numerical methods.

\subsection{Basic equations}
For a mass $M$ resting at the origin, the scalar fields are found to be given by
\begin{equation}\label{potfromTsol}
   \phi = - \frac{GM}{c^2} \, \frac{1}{r} ,
   \qquad \bar{\phi} = \psi = \bar{\psi} = 0 ,
\end{equation}
with $r=(x_1^2+x_2^2+x_3^2)^{1/2}$. The non-vanishing solution for the dimensionless scalar field $\phi$ is obtained from Eq.~(\ref{potfromT}) for $\Lambda=0$, a vanishing field $\bar{\phi}$ for static solutions follows from Eq.~(\ref{phievol}), and the conjugate momenta vanish according to the constraints (\ref{fundconstrs2}).

The coordinate condition (\ref{coco3mat2i}) for $\mu=0$ implies the concrete functional form
\begin{equation}\label{g0nstatiso}
   g_{0n} \propto \frac{x_n}{r^3} .
\end{equation}
In general relativity, $g_{0n}=0$ can be assumed without loss of generality because it can be achieved by a shift in time that depends on $r$ (see p.\,176 of \cite{Weinberg}). Moreover, this assumption is consistent with both quasi-Minkowskian and harmonic coordinate conditions, where the harmonic conditions are regarded as nearly Minkowskian (see, e.g., pp.\,163 and 254 of \cite{Weinberg}). As a decoupling of space and time is most natural for a static solution also in the composite theory of gravity defined on a background Minkowski space-time, we here assume a vanishing constant of proportionality in Eq.~(\ref{g0nstatiso}). This assumption actually leads to the most interesting static solutions. Therefore, we assume the following form of a static isotropic metric (see, e.g., Eq.~(8.1.3) of \cite{Weinberg}),
\begin{equation}\label{isoxg}
   g_{\mu\nu} = \left( \begin{matrix}
   -\beta & 0 \\
   0 & \alpha \, \delta_{mn} + \xi \, \frac{x_m x_n}{r^2}
   \end{matrix} \right) ,
\end{equation}
with inverse
\begin{equation}\label{isoxginv}
   \bar{g}^{\mu\nu} = \left( \begin{matrix}
   -\frac{1}{\beta} & 0 \\
   0 & \frac{\delta_{mn}}{\alpha} - \frac{\xi}{\alpha(\alpha+\xi)} \, \frac{x_m x_n}{r^2}
   \end{matrix} \right) ,
\end{equation}
where $\alpha$, $\beta$ and $\xi$ are dimensionless functions of $r$.

For the metric (\ref{isoxg}), the remaining coordinate conditions (\ref{coco3mat2i}) are given by the first-order ordinary differential equation
\begin{equation}\label{coco3matiso}
   \alpha' + \beta' - \xi' = \frac{4\xi}{r} - \frac{2GM}{c^2 r^2} ,
\end{equation}
where primes indicate derivatives with respect to $r$. Two further differential equations are obtained from the field equations (\ref{compactgeq}). Each equation $\Xi_{mn\mu'}=0$ leads to the same equation involving the third derivative of $\alpha$ when $\mu'$ equals $m$ or $n$,
\begin{eqnarray}\label{feqalpha}
   && 4 \alpha^2 (\alpha+\xi)^2 \beta r^3 \alpha^{(3)} + 3 \beta  \xi^2 r^3 \alpha'^3 \\
   && - \alpha  \beta  \xi  \alpha' r^2 \Big\{ 6 \xi r \alpha''
      - \alpha' ( 7 r \alpha'+2 r \xi' - 6 \xi) \Big\} \nonumber\\
   && + \alpha^2 \Big\{ \beta r \Big[ 2 \xi r \Big( \alpha'' ( 6 \xi - 2r \xi' - 9r \alpha' )
      - \alpha' ( r \xi''+13 \alpha' \nonumber\\
   && +4 \xi' ) \Big) + r^2 \alpha' (8 \alpha' \xi'+7 \alpha'^2+3 \xi'^2) + 4 \xi^2 \alpha' \Big]
      - \xi r^3 \alpha' \beta'^2
       \Big\} \nonumber\\
   && - \alpha^3 \Big\{ 2 \beta  \Big[ 2 r^2 \alpha'' ( 3 r \alpha' + r \xi' - 5 \xi)
      + r^2 \xi'' (r \alpha' + 2 \xi ) - 4 \xi^2 \nonumber\\
   && + r \alpha' ( 7 r \alpha' - 2 r \xi' + 2 \xi) - 3 r^2 \xi'^2 \Big]
      +  r^2 \beta'^2 \left(r \alpha'+2 \xi \right) \Big\} \nonumber\\
   && + 2 \alpha^4 \Big\{ 2 \beta ( 2 r^2 \alpha''- r^2 \xi'' -2 r \alpha' + 2 \xi )
      - r^2 \beta'^2 \Big\} = 0 . \nonumber
\end{eqnarray}
Finally, an equation involving the third derivative of $\beta$ is obtained from any of the field equations $\Xi_{0n0}=0$,
\begin{eqnarray}\label{feqbeta}
   && 4 \alpha (\alpha+\xi)^2 \beta^2 r^2 \beta^{(3)} - 2 \beta^2 \xi r^2 \alpha'^2 \beta' \\
   && + \alpha r \Big\{2 \beta \xi \beta' \big[ r \beta' (\alpha'+\xi')
      - \xi (3 r \beta''+2 \beta') \big] \nonumber\\
   && +\beta ^2 \Big[ r \beta' \left(6 \alpha' \xi'+\alpha'^2+3 \xi'^2\right)
      -2 \xi  \big[ \beta' ( r \alpha''+r \xi''+2 \xi' ) \nonumber\\
   && +2 \alpha' (r \beta''+3 \beta')+2 r \beta'' \xi' \big]
      +8 \xi^2 \beta'' \Big] +3 r \xi ^2 \beta'^3 \Big\} \nonumber\\
   && + 2 \alpha^2 \Big\{3 r^2 \xi  \beta'^3 + r \beta  \beta' \big[ r \beta' (\alpha '+\xi')
      \nonumber\\
   && - 2 \xi (3 r \beta''+2 \beta' ) \big]
      - \beta^2 \Big[ r \big[ \beta' ( r \alpha''+r \xi''+2 \xi' ) \nonumber\\
   && + 2 \alpha' (r \beta''+3 \beta')+2 r \beta'' \xi' \big]
      - 4 \xi \big[ 2 r \beta'' - \beta' \big] \Big] \Big\} \nonumber\\
   && + \alpha^3 \Big\{ 2\beta r \beta'' (4 \beta-3 r \beta') - 8 \beta^2 \beta'
      - 4 r \beta \beta'^2 + 3 r^2 \beta'^3 \Big\} = 0 . \nonumber
\end{eqnarray}
Note that these fairly complicated field equations contain only up to second derivatives of $\xi$. Equation (\ref{feqalpha}) contains third derivatives of $\alpha$, but only up to first derivatives of $\beta$, whereas Eq.~(\ref{feqbeta}) contains third derivatives of $\beta$ and up to second derivatives of $\alpha$.

The coordinate condition (\ref{coco3matiso}) and the field equations (\ref{feqalpha}), (\ref{feqbeta}) characterize the static isotropic metric (\ref{isoxg}) for composite gravity. In the following, we solve this system of three ordinary differential equations for $\alpha$, $\beta$ and $\xi$ by Robertson expansions in $1/r$ and by numerical methods. However, it is instructive and helpful to look also at the Yang-Mills variables for the static isotropic solution before proceeding to explicit solutions, that is, to consider the deeper structure of the composite theory, which can hardly be recognized in the complicated field equations (\ref{feqalpha}) and (\ref{feqbeta}).

The isotropic form of the gauge vector fields has previously been discussed in Appendix~C of \cite{hco240},
\begin{equation}\label{Aisotropicelegmatr}
   A^a_\mu = Y(r) \left(  \begin{matrix}
            x_1 & x_2 & x_3 & 0 & 0 & 0 \\
            0 & 0 & 0 & 0 & -x_3 & x_2 \\
            0 & 0 & 0 & x_3 & 0 & -x_1 \\
            0 & 0 & 0 & -x_2 & x_1 & 0 \\
            \end{matrix} \right) ,
\end{equation}
where the columns are labeled by the Lie algebra index $a$ and the rows by the space-time index $\mu$. Note that the Lorenz gauge condition (\ref{Lorenzgauge}) is satisfied for any differentiable function $Y(r)$. The property $A^{(kl)}_0=0$ of the static isotropic solution (\ref{Aisotropicelegmatr}) should be compared to the additional gauge conditions (\ref{gaugeconstrAa01}). A closed-form expression for $Y$ leading to a solution of the Yang-Mills field equations has been found in \cite{hco231,hco240},
\begin{equation}\label{YMexactsolYZ}
   Y = \frac{r_0}{r^2( r_0 +r )} ,
\end{equation}
where $r_0$ has been identified as $r_0=GM/c^2$. Note that $Y$ is a smooth function for all $r>0$ and has a $r^{-2}$ singularity at $r=0$.

In Appendix~C of \cite{hco240} it has been shown how this exact solution can be used to simplify the equations for the metric to a first-order system of ordinary differential equations. One can derive the following equations by post-processing of the exact solution for the Yang-Mills fields,
\begin{equation}\label{firstorderal}
   \alpha' = \frac{2\sqrt{(\alpha+\xi)\alpha}}{r_0+r} - 2 \frac{\alpha}{r} ,
\end{equation}
\begin{equation}\label{firstorderbe}
   \beta' = \frac{2r_0\sqrt{(\alpha+\xi)\beta}}{r(r_0+r)} ,
\end{equation}
\begin{equation}\label{firstorderxi}
   \xi' = \frac{2\sqrt{\alpha+\xi}}{r_0+r}
   \left( \sqrt{\alpha} + \frac{r_0}{r} \sqrt{\beta} \right)
   - 2 \frac{\alpha+2\xi}{r} + \frac{2 r_0}{r^2} .
\end{equation}
The first order system (\ref{firstorderal})--(\ref{firstorderxi}) is strikingly simpler than the system consisting of the coordinate condition (\ref{coco3matiso}) and the field equations (\ref{feqalpha}), (\ref{feqbeta}). The occurrence of square roots (which come with possible sign issues) is a consequence of the decomposition (\ref{localinertialdecomp}) of the metric into tetrad variables, where a symmetric decomposition has been found to be consistent with the gauge constraints verified after Eq.~(\ref{Aisotropicelegmatr}). The symmetric tetrad decomposition actually motivates the following change of variables,
\begin{equation}\label{isobstyle}
   \alpha = a^2 , \qquad \beta = b^2, \qquad \alpha+\xi = q^2 .
\end{equation}
With this transformation, the first-order system (\ref{firstorderal})--(\ref{firstorderxi}) further simplifies to a system without any square roots,
\begin{equation}\label{firstordera}
   a' = \frac{q}{r_0+r} - \frac{a}{r} ,
\end{equation}
\begin{equation}\label{firstorderb}
   b' = \frac{r_0 q}{r(r_0+r)} ,
\end{equation}
\begin{equation}\label{firstorderq}
   q' = \frac{2r a + r_0 b}{r(r_0+r)} - \frac{2q}{r} + \frac{r_0}{r^2 q} .
\end{equation}
After solving this compact first-order system of three ordinary differential equations, the transformation (\ref{isobstyle}) provides the static isotropic metric (\ref{isoxg}). The differential equations (\ref{firstordera})--(\ref{firstorderq}) imply the following singularities at $r=0$,
\begin{equation}\label{abqsingat0}
   a = a_0 \frac{r_0}{r} , \qquad b=-2 \sqrt{\frac{2 r_0}{7 r}} ,
   \qquad q=\sqrt{\frac{2 r_0}{7 r}} ,
\end{equation}
where the coefficient $a_0$ remains undetermined. The asymptotic results for $b$ and $q$ do not contain any free parameters.

\subsection{Robertson expansions}\label{secRobexpan}
Expansions of the isotropic metric in terms of $1/r$ are known as Robertson expansions. We insert the expansion
\begin{equation}\label{alphaRobertgen}
   \alpha(r) = \alpha_0 + \alpha_1 \frac{r_0}{r} + \alpha_2 \frac{r_0^2}{r^2}
   + \alpha_3 \frac{r_0^3}{r^3} + \alpha_4 \frac{r_0^4}{r^4} + \alpha_5 \frac{r_0^5}{r^5}
   + \ldots ,
\end{equation}
and similar expansions for $\beta(r)$ and $\xi(r)$ into Eqs.~(\ref{coco3matiso})--(\ref{feqbeta}). Convergence to the Minkoswki metric for $r \rightarrow \infty$ requires
\begin{equation}\label{Robertson0}
   \alpha_0 = \beta_0 = 1 , \qquad \xi_0 = 0 .
\end{equation}
The coefficients $\alpha_1$ and $\beta_1$ must be related by the cubic equation
\begin{eqnarray}\label{Robertson1}
   14 \alpha_1^3 + 12 \alpha_1^2 -18 \alpha_1 - 7 \beta_1^3 - 24 \beta_1^2 + 12 \beta_1 &&
   \nonumber\\ && \hspace{-14em}
   - 18 \alpha_1^2 \beta_1 + 6 \alpha_1 \beta_1^2 - 6 \alpha_1 \beta_1 -16 = 0 . \qquad 
\end{eqnarray}
With one exception, all other coefficients are determined by Eqs.~(\ref{coco3matiso})--(\ref{feqbeta}). Only the coefficient $\alpha_3$ remains undetermined.

For every $\alpha_1$, there is only one real solution $\beta_1$. The general result for $\beta_1$ involves third roots, but there is a remarkably simple exception: for $\alpha_1=1$, we obtain $\beta_1=-2$ and the resulting fifth-order Robertson expansions are 
\begin{equation}
   \alpha = 1 + \frac{r_0}{r} + \frac{9}{4} \frac{r_0^2}{r^2} + \alpha_3 \frac{r_0^3}{r^3}
   + \frac{7-3\alpha_3}{6} \frac{r_0^4}{r^4} - \frac{55-81\alpha_3}{60} \frac{r_0^5}{r^5} ,
   \nonumber
\end{equation}
\begin{equation}\label{allRobertson}
   \beta = 1 - 2 \frac{r_0}{r} + \frac{r_0^2}{r^2} + \frac{2}{3} \frac{r_0^3}{r^3}
   - \frac{7-3\alpha_3}{6} \frac{r_0^4}{r^4} + \frac{9-11\alpha_3}{10} \frac{r_0^5}{r^5} ,
\end{equation}
\begin{equation}
   \xi = \frac{r_0}{r} - \frac{13}{4} \frac{r_0^2}{r^2} - (2+3\alpha_3) \frac{r_0^3}{r^3}
   - \frac{1+3\alpha_3}{6} \frac{r_0^4}{r^4} - \frac{1-15\alpha_3}{12} \frac{r_0^5}{r^5} .
   \nonumber
\end{equation}
These expansions are remarkable also because they reproduce the high-precision predictions of general relativity for the gravitational redshift of spectral lines from white dwarf stars, the deflection of light (or electromagnetic waves outside the visible spectrum) by the sun, and the anomalous precession of the perihelion of Mercury. This can be verified by switching to the ``standard'' form of the metric according to Sec.~8.1 of \cite{Weinberg} (the required conditions are $\beta_1=-2$, $\alpha_1+\xi_1=2$, and $\beta_2=\alpha_1$).

The Robertson expansions (\ref{allRobertson}) have been derived directly from the equations for the metric, that is, without using the simplifications resulting from the exact expression for the gauge vector fields. From the simpler first-order systems (\ref{firstorderal})--(\ref{firstorderxi}) or (\ref{firstordera})--(\ref{firstorderq}), the same Robertson expansions (\ref{allRobertson}) follow more directly. No choice for $\alpha_1$ or $\beta_1$ in Eq.~(\ref{Robertson1}) is left. The remaining freedom of choosing the coefficient $\alpha_3$ suggests that one could introduce another constraint on the metric.

\subsection{Numerical solutions}
For the numerical solution of the systems of differential equations characterizing the static isotropic solution we proceed from large $r$ to small $r$. The required initial values at some large $r$ are obtained from the fifth-order Robertson expansions found in Sec.~\ref{secRobexpan}. We typically choose the initial $r$ between $10\,r_0$ and $100\,r_0$, where the larger values of $r$ are required for the higher derivative systems, whereas smaller $r$ can be afforded for the first-order system.

\begin{figure}
\centerline{\includegraphics[width=7 cm]{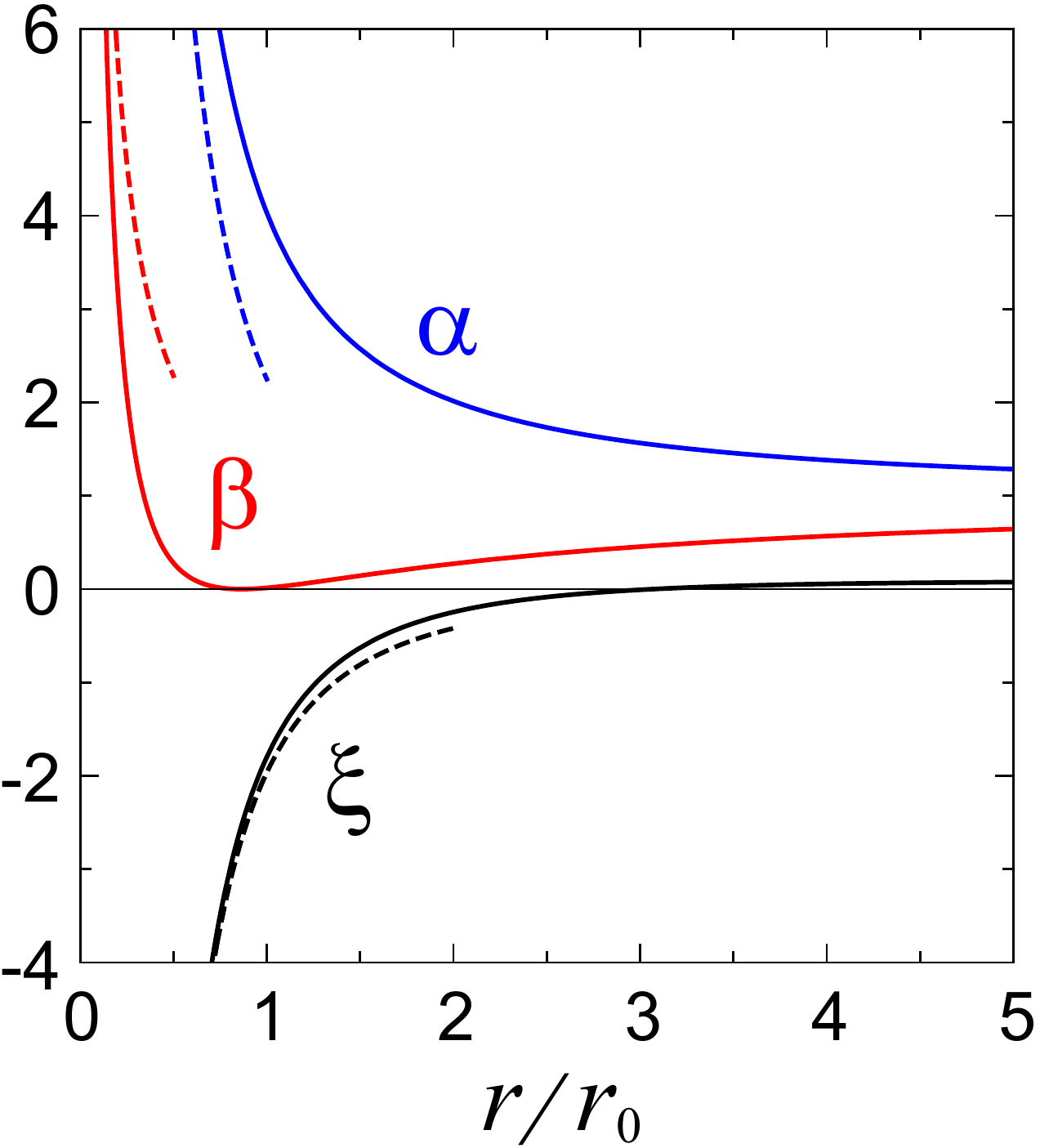}}
\caption[ ]{The functions $\alpha$, $\beta$ and $\xi$ characterizing the isotropic metric (\ref{isoxg}) obtained from the composite theory of gravity for $\alpha_3=-0.85$. The singular behavior (\ref{abqsingat0}) at $r=0$ is indicated by the respective dashed curves, where $a_0=1.5$ has been used as a fit parameter.} \label{complete_composite_gravity_fig_isotropic}
\end{figure}

As many of the equations in this section have been developed with the help of the {{\small MATLAB}\,\textregistered} software for symbolic mathematical computations, we have performed also the numerical integrations by means of the routine {\tt NDSolve} of {{\small MATLAB}\,\textregistered}. The first-order system (\ref{firstordera})--(\ref{firstorderq}) can be integrated without any problems. The resulting functions $\alpha(r)$, $\beta(r)$ and $\xi(r)$ depend on the choice of the free parameter $\alpha_3$ in the Robertson expansions. In general, the integration fails at some finite $r$. A particular value of $\alpha_3$ around $-0.85$ is required so that the integration can be continued al the way to the singular regime (\ref{abqsingat0}) around $r=0$. The solution that exists for all $r > 0$ is shown in Figure~\ref{complete_composite_gravity_fig_isotropic} together with the corresponding singularities (\ref{abqsingat0}) of $\alpha$, $\beta$ and $\xi$ at $r=0$.

The most interesting feature of the static isotropic solution of composite gravity is that the temporal component $\beta$ is always positive, except for a particular value of $r$ around $0.86\,r_0$, at which the $\beta$ curve reaches its minimum value zero. At this point, proper time does not proceed with the time in the background Minkowski system. This is the only singular feature of the black-hole solution around the Schwarzschild radius. Unlike in general relativity, we are not dealing with formal coordinate singularities in composite gravity so that any singular feature gains physical relevance and should hence be more gentle.

Integrating the third-order system (\ref{coco3matiso})--(\ref{feqbeta}) is considerably more challenging than solving first-order equations. Starting at large $r$, the integration fails when $\beta$ gets close to zero, even for $\alpha_3$ around $-0.85$. We then choose a point close to the minimum of $\beta$ where we still have reliable results and construct a fifth-order Taylor expansion around that point in the same way as we constructed a Robertson expansion around infinity by solving the field equations. This Taylor expansion is then used to get around the minimum of $\beta$ and to produce the initial values for continuing the numerical integration to smaller $r$. With this procedure, we can obtain exactly the same curves as shown in Figure~\ref{complete_composite_gravity_fig_isotropic}.  Again we find that $\alpha_3 \approx -0.85$ is required to obtain a smooth solution down to $r=0$.

\section{Summary, conclusions and outlook}
Every mathematical representation of nature comes with limitations for its range of validity. For general relativity, $90$ years of unshakable resistance to quantization suggest that this theory encounters limitations at small length scales. It is appealing to use a Yang-Mills theory for developing a theory of gravity that can be valid at short distances because all other known interactions between elementary particles are successfully modeled by this class of gauge theories---where the Lorentz group is the most natural symmetry group for a theory of gravity. The framework of composite theories allows us to select a small subset of physical solutions of the Yang-Mills theory with Lorentz symmetry group. What was missing in previous work \cite{hco231,hco240} was the proper characterization of the background Minkowski systems in which the composite theory of gravity should be valid.

The key advancement of this paper is the formulation of refined coordinate conditions characterizing the appropriate Minkowski systems for composite gravity. The third-order formulation (\ref{coco3mat}) shows that these coordinate conditions introduce a natural coupling of the gravitation field to the energy-momentum tensor of matter. The essence of that coupling is further highlighted in the scalar second-order differential equation (\ref{coco3mati}), which is accompanied by the homogeneous third-order equations (\ref{coco3mat0}).

The Hamiltonian formulation of the refined coordinate conditions is prepared by Eqs.~(\ref{coco3mat2i}) and (\ref{potfromT}): a first-order differential equation couples the metric to a scalar field, the d'Alembertian of which is given by the energy-momentum tensor. In other words, the interaction between gravitational field and matter is mediated by an additional scalar field. This interesting feature is a consequence of the higher derivative nature of the composite theory of gravity, which actually requires even two scalar fields and their conjugate momenta.

The refined coordinate conditions supplement the composite theory of gravity, which is obtained by expressing the gauge vector fields of the Yang-Mills theory with Lorentz symmetry group in terms of tetrad variables, in an ideal manner. In contrast to what we found in previous work on composite gravity \cite{hco231,hco240}, the refined coordinate conditions allow us to choose the coupling constant $\tilde{g}$ of the Yang-Mills theory as unity without loosing the famous predictions of general relativity tested with high precision. On the contrary, these predictions arise most naturally for $\tilde{g}=1$. Moreover, the connection (\ref{Gammadef}) associated with the gauge vector fields of the Yang-Mills theory is torsion-free, just as in general relativity.

The static isotropic solution of composite gravity does not only produce the proper field at distances much larger than the Schwarzschild radius, which determines the outcome of the high-precision tests, but also only mildly singular behavior around the Schwarzschild radius that characterizes the behavior of black holes. A previously observed characteristic of the black hole solutions of composite gravity \cite{hco231} is that at one particular distance from the center, which is of the order of the Schwarzschild radius, proper time does not grow with increasing time in the background Minkowski system (proper time for a fixed observer at that distance stands still).

Another interesting feature of the static isotropic solution is that there is actually a one-parameter family of solutions, at least at large distances $r$ from the center. Only one of these solutions is found to be smooth for all $r>0$, with power-law singularity at $r=0$. A general constraint for selecting the smooth solution might exist, which would change the count of constraints obtained from the canonical Hamiltonian formulation of composite gravity. The canonical count of six degrees of freedom resulting from $78$ constraints for $84$ variables (see Sec.~\ref{secdof}) might be further reduced by one or four degrees of freedom, depending on whether the additional constraint is scalar or vectorial.

The constraints for the conjugate momenta of the scalar variables occurring in the formulation of the coordinate conditions render possible geodesic particle motion. This is quite surprising because we have merely implemented an anisotropic velocity-momentum relation into a theory formulated in a background Minkowski space-time.

The canonical Hamiltonian formulation shows that we deal with two types of constraints: (i) constraints resulting from the composition rule and (ii) gauge constraints. As the former are second class constraints that can be handled by Dirac brackets \cite{Dirac50,Dirac58a,Dirac58b} and the latter can be treated by the BRST methodology \cite{BecchiRouetStora76,Tyutin75,Nemeschanskyetal86,hco229}, the path to the quantization of the composite theory of gravity seems to be clear. This is a major advantage of an approach starting from the class of Yang-Mills theories, which so successfully describe electro-weak and strong interactions and for which quantization is perfectly understood, and imposing Dirac-type constraints. For that reason, also the constraint(s) for selecting a smooth static isotropic solution should belong to Dirac's second class of constraints.

We have considered the coupling of the gravitational field to matter only for point particles. For many applications in astrophysics and cosmology, it would be important to find the proper coupling of gravity and fluid dynamics. Even in special relativity, the development of a theory of fluid dynamics has been a lengthy and thorny process, from pioneering work of Eckart \cite{Eckart40} and Landau and Lifshitz (see Chap.~XV of the textbook \cite{LandauLifshitz6}) to the widely used second-order theory of Israel and Stewart \cite{Israel76,IsraelStewart79} almost four decades later. Many authors contributed to the identification and solution of the problems of causality and stability, as well as the relations between these problems (see, for example, \cite{HiscockLindblom83,HiscockLindblom85,KostadtLiu00,VanBiro08,Denicoletal08,PuKoideRischke10}).

Within a thermodynamic framework, a special relativistic theory of fluid dynamics can be developed in a guided and robust manner \cite{hco109,hco111,hcobet} (for a more educational thermodynamic derivation, see Sec.~12.6 of \cite{hcomctp}). In particular, generating dissipative dynamics by entropy leads to stability for the entire range of physically meaningful model parameters \cite{hco234}. As these theories require more fields than the densities of mass, momentum and energy associated with universal conservation laws in nonrelativistic hydrodynamics, one should expect that the fluid dynamics of gases is different from the theory of liquids because the respective dissipation mechanisms are fundamentally different. In composite gravity, the introduction of gravitational interactions into fluid dynamics should happen through the tensorial character of mass or the anisotropy of the velocity-momentum relation in an undeformed space-time, as this anisotropy has been identified as the deeper reason for the occurrence of a metric. As the black holes come with a form of entropy \cite{Bekenstein73,BardeenCarHawk73,Hawking74,HawkingHunter99}, a thermodynamic framework for coupling gravitation and fluid dynamics \cite{hcobet,hco112,hco180,Jacobson95,Elingetal06} based on a non-canonical Hamiltonian formulation of reversible dynamics should be ideal.

\begin{acknowledgments}
I am grateful for the opportunity to do most of this work during my sabbatical year at the \emph{Collegium Helveticum} in Z\"urich.
\end{acknowledgments}

\appendix

\section{Proof of relation between covariant derivatives}\label{Appcovdevs}
The reformulation of equations for the Yang-Mills theory based on the Lorentz group in the metric language is based on the identity
\begin{equation}\label{supauxf1}
   f_{(\kappa\lambda)}^{bc} B_b C_c =
   \eta^{\kappa'\lambda'} \Big[ B_{(\kappa'\lambda)} C_{(\kappa\lambda')}
   - C_{(\kappa'\lambda)} B_{(\kappa\lambda')} \Big] ,
\end{equation}
which, in view of the definition (\ref{Btildef}), can be rewritten in the alternative form
\begin{equation}\label{supauxf2}
   {b^\kappa}_\mu {b^\lambda}_\nu f_{(\kappa\lambda)}^{bc} B_b C_c =
   \bar{g}^{\rho\sigma} \Big( \tilde{B}_{\rho\mu} \tilde{C}_{\sigma\nu}
   + \tilde{B}_{\rho\nu} \tilde{C}_{\mu\sigma} \Big) .
\end{equation}
These remarkably simple identities follow from the form of the structure constants of the Lorentz group. After writing the structure constants in the following explicit form (see Table~\ref{tabindexmatch} for the index conventions),
\begin{eqnarray}
   f^{abc} &=& \eta^{\kappa_a \lambda_c} \eta^{\kappa_b \lambda_a} \eta^{\kappa_c \lambda_b}
   - \eta^{\kappa_a \lambda_b} \eta^{\kappa_b \lambda_c} \eta^{\kappa_c \lambda_a}
   \nonumber\\
   &+& \eta^{\kappa_a \kappa_b} \big( \eta^{\kappa_c \lambda_a} \eta^{\lambda_b \lambda_c}
   - \eta^{\kappa_c \lambda_b} \eta^{\lambda_a \lambda_c} \big)
   \nonumber\\
   &+& \eta^{\kappa_a \kappa_c} \big( \eta^{\kappa_b \lambda_c} \eta^{\lambda_a \lambda_b}
   - \eta^{\kappa_b \lambda_a} \eta^{\lambda_b \lambda_c} \big)
   \nonumber\\
   &+& \eta^{\kappa_b \kappa_c} \big( \eta^{\kappa_a \lambda_b} \eta^{\lambda_a \lambda_c}
   - \eta^{\kappa_a \lambda_c} \eta^{\lambda_a \lambda_b} \big) ,
\label{Lorentzstructure}
\end{eqnarray}
the result (\ref{supauxf1}) is obtained by straightforward calculation.

We can now use Eq.~(\ref{supauxf2}) to evaluate the right-hand side of Eq.~(\ref{central}),
\begin{eqnarray}
   {b^\kappa}_\mu {b^\lambda}_\nu \left[
   \frac{\partial B_{(\kappa\lambda)}}{\partial x^\rho}
   + \tilde{g} \, f_{(\kappa\lambda)}^{bc} A_{b\rho} B_c \right] &=&
   \frac{\partial \tilde{B}_{\mu\nu}}{\partial x^\rho} \nonumber\\
   && \hspace{-12em} + \, \bar{g}^{\rho'\sigma} \bigg[ 
   \left( \tilde{g} \tilde{A}_{\rho'\mu\rho} - {b^\kappa}_{\rho'} \,
   \frac{\partial b_{\kappa\mu}}{\partial x^\rho} \right) \tilde{B}_{\sigma\nu}
   \nonumber\\ && \hspace{-9em} + \, 
   \left( \tilde{g} \tilde{A}_{\rho'\nu\rho} - {b^\kappa}_{\rho'} \,
   \frac{\partial b_{\kappa\nu}}{\partial x^\rho} \right) \tilde{B}_{\mu\sigma}
   \bigg] . \qquad 
\label{appcentral}
\end{eqnarray}
By using the expression (\ref{Atildefx}) and the definition (\ref{Gammadef}), we arrive at the fundamental relationship (\ref{central}) between the covariant derivatives associated with connections and the covariant derivatives associated with the Yang-Mills theory based on the Lorentz group.

\section{Alternative expression for field tensor}\label{Appfieldtensor}
From the definitions (\ref{Fdefinition}) and (\ref{Btildef}) and the fundamental relations (\ref{central}) and (\ref{supauxf2}), we obtain
\begin{eqnarray}
   \tilde{F}_{\mu\nu \mu'\nu'} &=& \frac{\partial \tilde{A}_{\mu\nu\nu'}}{\partial x^{\mu'}}
   - \Gamma^\sigma_{\mu'\mu} \tilde{A}_{\sigma\nu\nu'}
   + \Gamma^\sigma_{\mu'\nu} \tilde{A}_{\sigma\mu\nu'} \nonumber \\
   &-& \frac{\partial \tilde{A}_{\mu\nu\mu'}}{\partial x^{\nu'}}
   + \Gamma^\sigma_{\nu'\mu} \tilde{A}_{\sigma\nu\mu'}
   - \Gamma^\sigma_{\nu'\nu} \tilde{A}_{\sigma\mu\mu'} \nonumber \\
   &-& \tilde{g} \, \bar{g}^{\rho\sigma}
   \Big( \tilde{A}_{\rho\mu\mu'} \tilde{A}_{\sigma\nu\nu'}
   - \tilde{A}_{\rho\nu\mu'} \tilde{A}_{\sigma\mu\nu'} \Big) . \qquad 
\label{Fdefinitiontil}
\end{eqnarray}
By means of Eq.~(\ref{Atildefx}), we obtain
\begin{eqnarray}
   \tilde{g} \left( \frac{\partial \tilde{A}_{\mu\nu\nu'}}{\partial x^{\mu'}}
   - \frac{\partial \tilde{A}_{\mu\nu\mu'}}{\partial x^{\nu'}} \right) &=&
   \frac{\partial \tilde{\Gamma}_{\mu \mu' \nu}}{\partial x^{\nu'}}
   - \frac{\partial \tilde{\Gamma}_{\mu \nu' \nu}}{\partial x^{\mu'}} \nonumber\\
   && \hspace{-6em} + \, \frac{\partial {b^\kappa}_\mu}{\partial x^{\mu'}}
   \frac{\partial b_{\kappa\nu}}{\partial x^{\nu'}}
   - \frac{\partial {b^\kappa}_\mu}{\partial x^{\nu'}}
   \frac{\partial b_{\kappa\nu}}{\partial x^{\mu'}} ,
\label{Gammatildefxder}
\end{eqnarray}
and, again Eq.~(\ref{Atildefx}), gives
\begin{equation}\label{bderbder}
   \frac{\partial {b^\kappa}_\mu}{\partial x^{\mu'}}
   \frac{\partial b_{\kappa\nu}}{\partial x^{\nu'}} = \bar{g}^{\rho\sigma}
   ( \tilde{\Gamma}_{\rho\mu'\mu} + \tilde{g} \tilde{A}_{\rho\mu\mu'} )
   ( \tilde{\Gamma}_{\sigma\nu'\nu} + \tilde{g} \tilde{A}_{\sigma\nu\nu'} ) .
\end{equation}
By combining Eqs.~(\ref{Fdefinitiontil})--(\ref{bderbder}), we arrive at
\begin{equation}\label{FdefinitiontilGam}
   \tilde{F}_{\mu\nu \mu'\nu'} = \frac{1}{\tilde{g}} \bigg( 
   \frac{\partial \tilde{\Gamma}_{\mu \mu' \nu}}{\partial x^{\nu'}}
   - \frac{\partial \tilde{\Gamma}_{\mu \nu' \nu}}{\partial x^{\mu'}}
   + \tilde{\Gamma}_{\sigma \mu' \mu} \Gamma^\sigma_{\nu' \nu}
   - \tilde{\Gamma}_{\sigma \nu' \mu} \Gamma^\sigma_{\mu' \nu} \bigg) .
\end{equation}
This expression for the field tensor coincides with the one given in Eq.~(\ref{Fdefinitiontilz}) when the definition (\ref{Gammabardef}) of the connection is used.

\section{Classification of field equations}\label{Appclassfieldeqs}
The $24$ Yang-Mills field equations for the metric given in Eq.~(\ref{compactgeq}) are partially evolution equations and partially constraints. We here classify them according to the form of their third-derivative terms.

The first class of six equations is given by
\begin{eqnarray}
   {\cal D}_3 \big( \Xi_{0 n m'} + \Xi_{0 m' n} \big) &=&
   - \frac{\partial^3 g_{m'n}}{c^3 \partial t^3} \nonumber\\
   && \hspace{-9.5em} + \, \frac{\partial^2}{c^2 \partial t^2}
   \left( \frac{\partial g_{0n}}{\partial x^{m'}}
   + \frac{\partial g_{0m'}}{\partial x^n} \right) \nonumber\\
   && \hspace{-9.5em} + \, \frac{\partial}{c \partial t} \left( \Delta g_{m'n} -
   \frac{\partial^2 g_{00}}{\partial x^{m'} \partial x^n}
   - \frac{1}{2} \frac{\partial^2 g_{mn}}{\partial x^{m'} \partial x_m}
   - \frac{1}{2} \frac{\partial^2 g_{mm'}}{\partial x^n \partial x_m} \right) \nonumber\\
   && \hspace{-9.5em} + \, \frac{\partial^3 g_{0m}}{\partial x^{m'} \partial x^n \partial x_m}
   - \frac{1}{2} \Delta \left( \frac{\partial g_{0n}}{\partial x^{m'}}
   + \frac{\partial g_{0m'}}{\partial x^n} \right) ,
\label{geqclass1}
\end{eqnarray}
where ${\cal D}_3$ extracts the third-derivative terms from Eq.~(\ref{compactgeq}). The occurrence of third time derivatives demonstrates that this class of equations consists of the evolution equations for the spatial components $g_{m'n}$ of the metric.

For the next class of six equations, given by
\begin{eqnarray}
   {\cal D}_3 \big( \Xi_{m n m'} + \Xi_{m m' n} \big) &=&
   - \frac{1}{2} \square \left( \frac{\partial g_{mn}}{\partial x^{m'}}
   + \frac{\partial g_{mm'}}{\partial x^n} \right) ,
   \nonumber\\
   && \hspace{-9.5em} + \, \frac{\partial}{\partial x^m} \left( \square g_{m'n} 
   - \frac{1}{2} \frac{\partial^2 g_{n\rho}}{\partial x^{m'} \partial x_\rho}
   - \frac{1}{2} \frac{\partial^2 g_{m'\rho}}{\partial x^n \partial x_\rho} \right) \nonumber\\
   && \hspace{-9.5em} +
   \, \frac{\partial^3 g_{m\rho}}{\partial x^{m'} \partial x^n \partial x_\rho} ,
\label{geqclass2}
\end{eqnarray}
there are no third time derivatives. This is equally true for
\begin{eqnarray}
   {\cal D}_3 \big( \Xi_{\mu n m'} - \Xi_{\mu m' n} \big) &=&
   \frac{1}{2} \frac{\partial}{\partial x^n} \left(
   \frac{\partial^2 g_{m'\rho}}{\partial x^\mu \partial x_\rho} - \square g_{\mu m'} \right) 
   \nonumber\\
   && \hspace{-5em} - \, \frac{1}{2} \frac{\partial}{\partial x^{m'}} \left(
   \frac{\partial^2 g_{n\rho}}{\partial x^\mu \partial x_\rho} - \square g_{\mu n} \right) ,
\label{geqclass3}
\end{eqnarray}
where all six equations of this class contain at least one spatial derivative in their third derivative terms. The final six equations belong to the class with
\begin{eqnarray}
   {\cal D}_3 \big( \Xi_{\mu n 0} \big) &=&
   \frac{1}{2} \frac{\partial}{\partial x^n} \left(
   \frac{\partial^2 g_{\mu\rho}}{\partial x^0 \partial x_\rho} - \square g_{0 \mu} \right)
   \nonumber\\
   && \hspace{-3em} - \, \frac{1}{2} \frac{\partial}{\partial x^\mu} \left(
   \frac{\partial^2 g_{nm}}{\partial x^0 \partial x_m} - \Delta g_{0 n} \right) ,
\label{geqclass4}
\end{eqnarray}
where, even for $\mu=0$, at least one spatial derivative occurs in the third derivative terms. In conclusion, we have found six evolution equations for the spatial components $g_{m'n}$ of the metric and $18$ constraints.

\section{Conserved fluxes}\label{Appconsflux}
In the presence of external fluxes, Eq.~(\ref{YMfieldeqs}) implies that the alternative form (\ref{YMfieldeqsR}) of the Yang-Mills equations can be generalized to
\begin{equation}\label{YMfieldeqsRJ}
   \frac{\partial {R^\mu}_{\nu\mu'\nu'}}{\partial x_{\nu'}} +
   \eta^{\rho\nu'} \left( \Gamma^\mu_{\rho\sigma} {R^\sigma}_{\nu\mu'\nu'}
   - \Gamma^\sigma_{\rho\nu} {R^\mu}_{\sigma\mu'\nu'} \right) =
   \hat{J}^{\mu\sigma}_{\mu'} g_{\sigma\nu} .
\end{equation}
We then realize that the fluxes $\hat{J}^{\mu\nu}_{\mu'}$ must satisfy the local conservation law
\begin{equation}\label{Jhatconservation}
   \frac{\partial}{\partial x_{\mu'}} \Big( \hat{J}^{\mu\sigma}_{\mu'} g_{\sigma\nu} \Big) =
   \eta^{\rho\nu'} \frac{\partial}{\partial x_{\mu'}}
   \Big( \Gamma^\mu_{\rho\sigma} {R^\sigma}_{\nu\mu'\nu'}
   - \Gamma^\sigma_{\rho\nu} {R^\mu}_{\sigma\mu'\nu'} \Big) .
\end{equation}

Alternatively, after introducing the tensor
\begin{equation}\label{Rhatdef}
   \mbox{$\hat{R}^{\mu\nu}$}_{\mu'\nu'} = \bar{g}^{\nu\sigma} {R^\mu}_{\sigma\mu'\nu'} =
   \bar{g}^{\mu\rho} \bar{g}^{\nu\sigma} \tilde{F}_{\rho\sigma \mu'\nu'} ,
\end{equation}
we can rewrite Eq.~(\ref{YMfieldeqsRJ}) as
\begin{equation}\label{YMfieldeqsRJx}
   \frac{\partial \hat{R}^{\mu\nu\mu'\nu'}}{\partial x^{\nu'}}
   + \Gamma^\mu_{\nu'\sigma} \hat{R}^{\sigma\nu\mu'\nu'}
   + \Gamma^\nu_{\nu'\sigma} \hat{R}^{\mu\sigma\mu'\nu'} = \hat{J}^{\mu\nu\mu'} .
\end{equation}
We then obtain an alternative formulation of the local conservation law for the fluxes,
\begin{equation}\label{Jhatconservationx}
   \frac{\partial \hat{J}^{\mu\nu\mu'}}{\partial x^{\mu'}} =
   \frac{\partial}{\partial x^{\mu'}} \Big( \Gamma^\mu_{\nu'\sigma} \hat{R}^{\sigma\nu\mu'\nu'}
   + \Gamma^\nu_{\nu'\sigma} \hat{R}^{\mu\sigma\mu'\nu'} \Big) .
\end{equation}

\section{Some derivatives}\label{Appsomeder}
For fixed particle positions and metric, Eq.~(\ref{particleppm}) implies
\begin{equation}\label{dHmdp1}
   \bar{g}^{\mu 0} p_\mu \delta p_0 + \bar{g}^{\mu j} p_\mu \delta p_j = 0 ,
\end{equation}
which, with $H_{\rm m}=-c p_0$, leads to the partial derivative
\begin{equation}\label{dHmdp2}
   \frac{\partial H_{\rm m}}{\partial p_j}
   = \frac{1}{\gamma m} \, \bar{g}^{j\mu} p_\mu .
\end{equation}
For fixed particle positions and momenta, Eq.~(\ref{particleppm}) implies
\begin{equation}\label{dHmdg1}
   \frac{1}{2} \gamma m \frac{d\bar{x}^\mu}{d t} \frac{d\bar{x}^\nu}{d t} \,
   \delta g_{\mu\nu} = c \, \delta p_0 = - \delta H_{\rm m} ,
\end{equation}
which leads to the functional derivative
\begin{equation}\label{dHmdg2}
   \bar{b}^{\mu\kappa} \frac{\delta H_{\rm m}}{\delta {b^\kappa}_\nu} =
   - \gamma m \frac{d\bar{x}^\mu}{d t} \frac{d\bar{x}^\nu}{d t}
   \, \delta^3(\bm{x}-\bar{\bm{x}}(t)) =
   - \bar{g}^{\mu\rho} \, {T_\rho}^\nu .
\end{equation}
This functional derivative is symmetric in $\mu$ and $\nu$.

From Eqs.~(\ref{velmomrel}) and (\ref{gammadef}) we obtain
\begin{equation}\label{dgamdg1}
   - \frac{m c^2}{\gamma} = p_\mu \frac{d\bar{x}^\mu}{d t} =
   c p_0 + \frac{\bar{g}^{n \mu} p_n p_\mu}{\gamma m} ,
\end{equation}
which we again differentiate for fixed particle positions and momenta. After some rearrangements, we obtain
\begin{equation}\label{dgamdg2}
   p_0 \, \gamma \delta \left( \frac{1}{\gamma} \right) =
   \left( 2 - \frac{\bar{g}^{00} p_0}{\gamma m c} \right) \delta p_0
   + \frac{p_n p_\mu}{\gamma m c} \, \delta \bar{g}^{n \mu} ,
\end{equation}
which, by means of Eq.~(\ref{dHmdg1}) and the differential of inverse matrices, leads to the final result
\begin{equation}\label{dgamdg3}
   \delta \left( \frac{1}{\gamma} \right) = \frac{1}{2\gamma}
   \left( \bar{g}^{0\mu} \frac{d\bar{x}^\nu}{d x^0}
   + \bar{g}^{0\nu} \frac{d\bar{x}^\mu}{d x^0}
   - \bar{g}^{00} \frac{d\bar{x}^\mu}{d x^0} \frac{d\bar{x}^\nu}{d x^0} \right)
   \delta g_{\mu\nu} .
\end{equation}
This result can be rewritten in the alternative form
\begin{equation}\label{dgamdg4}
   \bar{b}^{\mu\kappa} \frac{\delta}{\delta {b^\kappa}_\nu} \frac{mc^2}{\gamma} =
   \frac{1}{2\gamma^2} \left( \bar{g}^{0\mu} \hat{T}^{0\nu}
   + \bar{g}^{0\nu} \hat{T}^{0\mu} - \bar{g}^{00} \hat{T}^{\mu\nu} \right) .
\end{equation}

\section{Further evolution equations and identities}
From Eqs.~(\ref{Aevolaj}) and (\ref{Atildefx}) for $\tilde{g}=1$, we obtain the following evolution equations for the connection variables,
\begin{equation}\label{Gamtilevol}
   \frac{\partial \tilde{\Gamma}_{\mu m \nu}}{\partial x^0} = \tilde{E}_{\mu\nu m}
   + \frac{\partial \tilde{\Gamma}_{\mu\nu 0}}{\partial x^m}
   - \Gamma^\sigma_{\mu m} \tilde{\Gamma}_{\sigma 0\nu}
   + \Gamma^\sigma_{\nu m} \tilde{\Gamma}_{\sigma 0\mu} ,
\end{equation}
which may also be regarded as the identification of $\tilde{E}_{\mu\nu m}$ as one of the components of Eq.~(\ref{FdefinitiontilGam}). From Eq.~(\ref{R4def}), we similarly obtain
\begin{equation}\label{Gamevol}
   \frac{\partial \Gamma^\mu_{m \nu}}{\partial x^0} = {R^\mu}_{\nu m 0}
   + \frac{\partial \Gamma^\mu_{0 \nu}}{\partial x^m}
   + \Gamma^\mu_{m \sigma} \Gamma^\sigma_{0\nu}
   - \Gamma^\sigma_{m \nu} \Gamma^\mu_{0 \sigma} .
\end{equation}
From the representation $\tilde{\Gamma}_{\mu 00} = X_{\mu 0} - \tilde{A}_{\mu 00}$ (see Eqs.~(\ref{Atildefx}) and (\ref{bevol})), we find
\begin{equation}\label{Gamtil00evol}
   \frac{\partial \tilde{\Gamma}_{n00}}{\partial x^0} =
   \frac{\partial}{\partial x_l} \left( \frac{\partial g_{0n}}{\partial x^l}
   - 2 \tilde{\Gamma}_{0nl} \right) 
   + \frac{\partial}{\partial x^n} \big( \tilde{\Gamma}_{0ll} - \bar{\phi} \big) ,
\end{equation}
which can be further simplified by means of the following identity implied by the coordinate condition (\ref{coco3mat2i}),
\begin{equation}\label{phibaralt}
   \bar{\phi} = \frac{\partial \phi}{\partial x^0} =
   \eta^{\mu\nu} \, \tilde{\Gamma}_{0\mu\nu} =
   \tilde{\Gamma}_{0ll} - \tilde{\Gamma}_{000} .
\end{equation}
This latter identity moreover allows us to calculate the time derivative of $\tilde{\Gamma}_{000}$ from Eqs.~(\ref{phibevol}) and (\ref{Gamtilevol}).

Evolution equations for $\tilde{E}_{\mu\nu j}$ are obtained from (\ref{Eevolaj}),
\begin{eqnarray}
   \frac{\partial \tilde{E}_{\mu\nu j}}{\partial x^0} &=&
   - \frac{\partial \tilde{F}_{\mu\nu mj}}{\partial x_m}
   + \eta^{\mu'\mu''} \left( 
   \Gamma^\sigma_{\mu'\mu} \tilde{F}_{\sigma\nu \mu''j}
   + \Gamma^\sigma_{\mu'\nu} \tilde{F}_{\mu\sigma \mu''j} \right) \nonumber\\
   && - \tilde{J}_{\mu\nu j} ,
\label{Etilevol}
\end{eqnarray}
where the Lorenz gauge $E_{(\kappa\lambda) 0}=0$ and the definition
\begin{equation}\label{Jtildef}
   \tilde{J}_{\mu\nu \mu'} = 
   {b^\kappa}_\mu {b^\lambda}_\nu J_{(\kappa\lambda) \mu'} =
   g_{\mu\rho} g_{\nu\sigma} \hat{J}^{\rho\sigma}_{\mu'} ,
\end{equation}
have been used. The local conservation law (\ref{Jhatconservation}) becomes
\begin{equation}\label{Jtildeconservation}
   \frac{\partial}{\partial x_{\mu'}} \Big( \bar{g}^{\mu\sigma} \tilde{J}_{\sigma\nu\mu'} \Big) =
   \eta^{\rho\nu'} \frac{\partial}{\partial x_{\mu'}}
   \Big( \Gamma^\mu_{\rho\sigma} {R^\sigma}_{\nu\mu'\nu'}
   - \Gamma^\sigma_{\rho\nu} {R^\mu}_{\sigma\mu'\nu'} \Big) .
\end{equation}
Equation (\ref{Etilevol}) may be regarded as an evolution equation for $\tilde{F}_{\mu\nu 0j}$. We next derive an evolution equation for the components $\tilde{F}_{\mu\nu mn}$. From the definition (\ref{Fdefinition}) of the field tensor and the evolution equations (\ref{Aevolaj}), we obtain
\begin{eqnarray}
    \frac{\partial F_{a mn}}{\partial x^0} + f_a^{bc} \, A_{b0} F_{c mn} &=&
    \frac{\partial E_{a m}}{\partial x^n} + f_a^{bc} \, A_{bn} E_{c m} \nonumber\\
    &-& \boxed{m \leftrightarrow n} .
\label{Fspatialevol}
\end{eqnarray}
The identity (\ref{central}) leads to the desired result
\begin{eqnarray}
    \frac{\partial \tilde{F}_{\mu\nu mn}}{\partial x^0}
    - \Gamma^\sigma_{0\mu} \tilde{F}_{\sigma\nu mn}
    - \Gamma^\sigma_{0\nu} \tilde{F}_{\mu\sigma mn} &=& \nonumber\\
    && \hspace{-14.5em}
    \frac{\partial \tilde{E}_{\mu\nu m}}{\partial x^n}
    - \Gamma^\sigma_{n\mu} \tilde{E}_{\sigma\nu m}
    - \Gamma^\sigma_{n\nu} \tilde{E}_{\mu\sigma m} 
    - \boxed{m \leftrightarrow n} . \qquad \quad 
\label{Ftilspatialevol}
\end{eqnarray}

Equation (\ref{Aevolaj}) implies the representation (\ref{Fdefinitiontilz}) for $\tilde{F}_{\mu\nu j0}$. By direct calculation we then find the identity
\begin{equation}\label{FJacobi}
   \tilde{F}_{mn j0} + \tilde{F}_{jm n0} + \tilde{F}_{nj m0} = 0 ,
\end{equation}
which leads to
\begin{equation}\label{EJacobi}
   \tilde{E}_{mn j} + \tilde{E}_{jm n} + \tilde{E}_{nj m} = 0 .
\end{equation}
More generally, Eq.~(\ref{Fdefinitiontilz}) implies
\begin{equation}\label{FJacobix}
   \tilde{F}_{\mu\nu\rho\sigma} + \tilde{F}_{\nu\rho\mu\sigma}
   + \tilde{F}_{\rho\mu\nu\sigma} = 0 .
\end{equation}

Similarly, the representation (\ref{Fdefinitiontilz}) leads to the identity
\begin{equation}\label{EderJacobi}
   \frac{\partial \breve{E}_{mn j}}{\partial x^l} +
   \frac{\partial \breve{E}_{nl j}}{\partial x^m} +
   \frac{\partial \breve{E}_{lm j}}{\partial x^n} = 0 ,
\end{equation}
with
\begin{equation}\label{Ebrevedef}
   \breve{E}_{\mu\nu j} = \tilde{E}_{\mu\nu j}
   - \Gamma^\sigma_{j\mu} \tilde{\Gamma}_{\sigma 0\nu}
   + \Gamma^\sigma_{j\nu} \tilde{\Gamma}_{\sigma 0\mu} .
\end{equation}
As a direct consequence of the field equations, we obtain
\begin{equation}\label{Etildivergence}
   \frac{\partial \tilde{E}_{\mu\nu j}}{\partial x_j} =
   \Gamma^\sigma_{j\mu} \tilde{E}_{\sigma\nu j} +
   \Gamma^\sigma_{j\nu} \tilde{E}_{\mu\sigma j} - \tilde{J}_{\mu\nu 0} .
\end{equation}

\section{Representations in terms of metric}
From Eq.~(\ref{Fdefinitiontilz}) we obtain the representation
\begin{equation}\label{Etilgrep}
   \tilde{E}_{\mu\nu l} = \frac{1}{2} \frac{\partial}{\partial x^\mu}
   \left( \frac{\partial g_{0\nu}}{\partial x^l}
   - \frac{\partial g_{l\nu}}{\partial x^0} \right) 
   + \Gamma^\sigma_{\mu l} \tilde{\Gamma}_{\sigma 0\nu}
   - \; \boxed{\mu \leftrightarrow \nu} .
\end{equation}
The representation (\ref{Etilgrep}) implies that the secondary constraints (\ref{METqlinconstraints2a}) and (\ref{METqlinconstraints2b}), which express symmetry conditions, are satisfied identically. The tertiary constraints (\ref{METqlinconstraints3a}) express the field equations $\Xi_{mnj}=0$ given in Eq.~(\ref{compactgeq}), whereas the constraints (\ref{METqlinconstraints3b}) can be recognized as $\Xi_{0nm}-\Xi_{0mn}=0$. These $12$ conditions correspond to the constraints (\ref{geqclass2}) and (\ref{geqclass3}) identified in Appendix~\ref{Appclassfieldeqs}. As a consequence of the Lorenz gauge, the evolution equations for $E_{a0}$ result in the six constraints (\ref{geqclass4}).


%

\end{document}